\documentclass[prb,twocolumn]{revtex4}%

\usepackage{graphicx}
\usepackage{amsmath}
\usepackage{ulem}
\usepackage{color}

\newcommand{\be}{\begin{equation}}
\newcommand{\ee}{\end{equation}}
\newcommand{\bea}{\begin{eqnarray*}}
\newcommand{\eea}{\end{eqnarray*}}
\newcommand{\bean}{\begin{eqnarray}}
\newcommand{\eean}{\end{eqnarray}}

\begin{document}

\draft
\title{\bf Thermal rectification through the
topological states of asymmetrical length armchair graphene
nanoribbons heterostructures with vacancies}

\author{David M T Kuo}

\address{Department of Electrical Engineering and Department of Physics, National Central
University, Chungli, 32001 Taiwan, China}

\date{\today}

\begin{abstract}
We present a theoretical investigation of electron heat current in
asymmetrical length armchair graphene nanoribbon (AGNR)
heterostructures with vacancies, focusing on the topological
states (TSs). In particular, we examine the 9-7-9 AGNR
heterostructures where the TSs are well-isolated from the
conduction and valence subbands. This isolation effectively
mitigates thermal noise of subbands arising from temperature
fluctuations during charge transport. Moreover, when the TSs
exhibit an orbital off-set, intriguing electron heat rectification
phenomena are observed, primarily attributed to inter-TS electron
Coulomb interactions. To enhance the heat rectification ratio
($\eta_Q$), we manipulate the coupling strengths between the heat
sources and the TSs by introducing asymmetrical lengths in the
9-AGNRs. This approach offers control over the rectification
properties, enabling significant enhancements. Additionally, we
introduce vacancies strategically positioned between the heat
sources and the TSs to suppress phonon heat current. This
arrangement effectively reduces the overall phonon heat current,
while leaving the TSs unaffected. Our findings provide valuable
insights into the behavior of electron heat current in AGNR
heterostructures, highlighting the role of topological states,
inter-TS electron Coulomb interactions, and the impact of
structural modifications such as asymmetrical lengths and vacancy
positioning. These results pave the way for the design and
optimization of graphene-based devices with improved thermal
management and efficient control of electron heat transport.
\end{abstract}

\maketitle

\section{Introduction}
Thermal rectification, a widely studied phenomenon in condensed
matter physics, refers to the asymmetric heat flow within a
material or system, where heat preferentially conducts in one
direction over the other [\onlinecite{DubiY}-\onlinecite{ZhaoH}].
This property holds significant implications for various fields,
including thermal management, energy conversion, and nanoscale
heat transfer [\onlinecite{DubiY}-\onlinecite{ZhaoH}].
Consequently, considerable efforts have been devoted to developing
heat diodes (HDs) capable of rectifying heat flow
[\onlinecite{StarrC}-\onlinecite{WangY}]. These HDs consider
different types of heat carriers, such as phonons
[\onlinecite{Ter}-\onlinecite{ChenXK}], electrons
[\onlinecite{Kuo1}--\onlinecite{Vipul}], photons
[\onlinecite{Otey},\onlinecite{BasuS}] and other excitation modes
[\onlinecite{ZhangY},\onlinecite{YuanMQ}]. The rectification ratio
of HDs, denoted as $\eta_Q = Q_F/|Q_B| - 1$, quantifies the heat
currents in the forward ($Q_F$) and backward ($Q_B$) temperature
biases. An ideal HD exhibits an extremely high ratio of
$Q_F/|Q_B|$, indicating a nearly complete blockade of heat current
in the backward temperature bias. However, experimental findings
of the rectification ratio of thermoelectric HDs typically range
between 0.07 and 0.5 [\onlinecite{ChangCW}-\onlinecite{ZhangXK}],
falling short of the requirement for applications demanding a
rectification ratio greater than 10[\onlinecite{ChenXK}].
Nevertheless, in metal/superconductor junction systems operating
at extremely low temperatures ($T < 1 K$) authors demonstrated a
high-efficiency electron HD with $\eta_Q = 139$
[\onlinecite{Perez}]. The results of ref. [\onlinecite{Perez}]
showcased the significant suppression of phonon excitations at
temperatures below than liquid helium temperature.

Although 1-D systems are not ideal for realistic HD
applications[\onlinecite{Ter},\onlinecite{LiBW}], the mechanism of
phonon heat rectification has been clearly revealed
[\onlinecite{Ter},\onlinecite{LiBW}]. For practical applications,
some studies have extended 1-D systems to 2D materials, such as
graphene nanostructures, for phonon HDs
[\onlinecite{HuJN}-\onlinecite{WangY}]. These mechanisms typically
involve phonon transport through asymmetrical graphene
nanostructures coupled to hot and cold reservoirs. However,
synthesizing these phonon HD structures using current bottom-up
techniques remains challenging
[\onlinecite{Cai}--\onlinecite{SunQ}]. One potential solution lies
in the fabrication of armchair graphene nanoribbon (AGNR)
heterostructures, which can be achieved through bottom-up
synthesis techniques [\onlinecite{DRizzo}--\onlinecite{SunQ}]. The
introduction of defects or vacancies in AGNRs can strongly
suppress phonon heat currents
[\onlinecite{JWJiang}-\onlinecite{SHTan}]. Consequently, electron
heat current can significantly surpass phonon heat current at low
temperatures. Surprisingly, up to date, nonlinear electron and
heat transport in AGNR heterostructures have received limited
attention [\onlinecite{Jacobse}--\onlinecite{Kuo5}], particularly
regarding electron heat rectification behavior. Therefore, a
notable research gap exists in the nonlinear heat transport
properties of AGNR heterostructures that requires further
investigation.

The interface states of 9-7-9 AGNR heterostructures exhibit
topological states (TSs) with energy levels near the
charge-neutral point, which are well-separated from the conduction
and valence subbands [\onlinecite{DRizzo},\onlinecite{DJRizzo}].
This unique characteristic allows for the manipulation of electron
transport solely through TSs, avoiding the subband channels. The
strong intra-TS and inter-TS electron Coulomb interactions will
lead to remarkable current rectification and negative differential
conductance in charge transport through these serially coupled TSs
(SCTS) in the Pauli spin blockade (PSB) configuration, similar to
the behavior of serially coupled quantum dots [\onlinecite{Ono}].

This article aims to theoretically investigate the electron heat
rectification of SCTSs composed of asymmetrical length 9-7-9 AGNRs
with vacancies, as depicted in Figure 1(b) and 1(c). Asymmetrical
length 9-7-9 AGNR heterostructures offer tunable tunneling rates
for the left and right TSs, with the energy level of the right TS
being modulated by the right gate electrode (shown in Figure
1(b)). The vacancies effectively suppress phonon heat current,
while the TSs remain largely unaffected since their wave functions
are located at the interfaces between 9-AGNR and 7-AGNR (Fig.
1(a)). Consequently, electron heat current dominates over phonon
heat current in this scenario, leading to the observation of
asymmetrical electron heat current with high rectification
efficiency above liquid-helium temperature.

\begin{figure}[h]
\centering
\includegraphics[trim=2.5cm 0cm 2.5cm 0cm,clip,angle=0,scale=0.3]{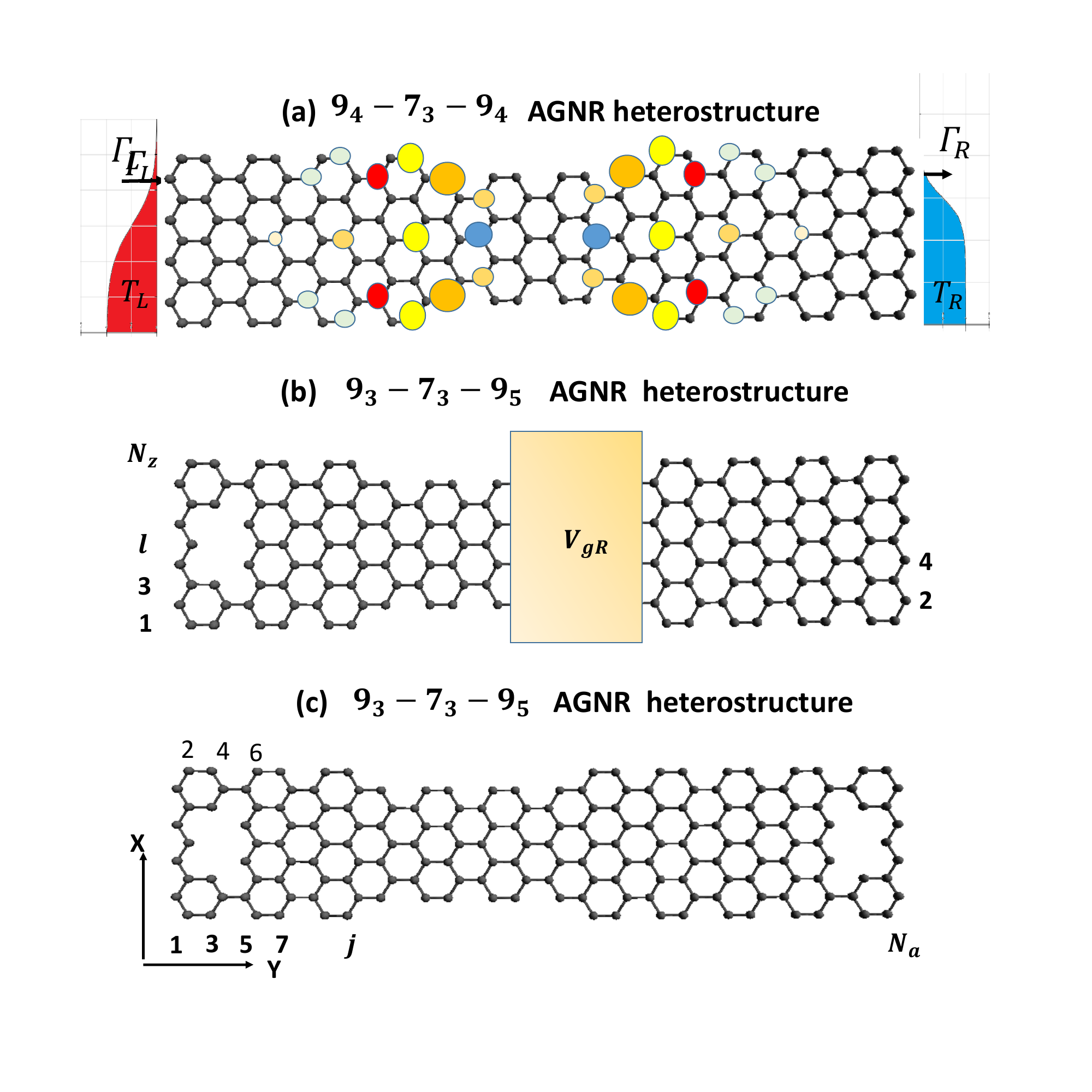}
\caption{(a) Schematic diagram of a symmetrical heterostructure
composed of $9_4-7_3-9_4$ AGNR connected to electrodes. The
tunneling rate of electrons between the left (right) electrode and
the leftmost (rightmost) atoms of the AGNR is denoted by
$\Gamma_{L}$ ($\Gamma_R$), respectively. The temperature of the
left ($L$) and right ($R$) electrodes is represented by $T_L$ and
$T_R$, respectively. The charge density of the topological state
with energy $\varepsilon = 0.1179$~eV is plotted for a
$9_4-7_3-9_4$ AGNR heterostructure. (b) and (c) Schematic diagrams
of asymmetrical length  AGNR heterostructures, namely
$9_3-7_3-9_5$, with one triple-vacancy and two triple-vacancies,
respectively. The subscripts $w$, $x$, and $y$ in the notation
$9_w-7_x-9_y$ denote the segment lengths of the AGNR
heterostructures in terms of the unit cell (u.c.).}
\end{figure}

\section{Calculation method}
We investigate the transport properties of heterostructures
composed of 9-7-9 AGNRs connected to electrodes, as shown in Fig.
1. To analyze this system, we employ a combination of the
tight-binding model and the Green's function technique
[\onlinecite{HaugH}--\onlinecite{Kuo3}]. The Hamiltonian of the
system consists of two parts: $H = H_0 + H_{\text{AGNR}}$. Here,
$H_0$ represents the Hamiltonian of the electrodes themselves,
including the coupling between the electrodes and the 9-7-9
AGNR[\onlinecite{Kuo2},\onlinecite{Kuo3}]. $H_{\text{AGNR}}$
corresponds to the Hamiltonian for the 9-7-9 AGNR
heterostructures, which can be expressed as follows:

\begin{small}
\begin{eqnarray}
H_{AGNR} &= &\sum_{\ell,j} E_{\ell,j}~d^{\dagger}_{\ell,j}d_{\ell,j}\\
\nonumber&-& \sum_{\ell,j}\sum_{\ell',j'} t_{(\ell,j),(\ell', j')}
d^{\dagger}_{\ell,j} d_{\ell',j'} + h.c,
\end{eqnarray}
\end{small}

In this equation, $E_{\ell,j}$ represents the on-site energy for
the $p_z$ orbital in the $\ell$-th row and $j$-th column. The
study neglects the spin-orbit interaction. The creation
(destruction) of an electron at the atom site labeled by ($\ell$,
$j$), where $\ell$ and $j$ are the row and column indices,
respectively, is denoted by $d^{\dagger}_{\ell,j}$ ($d_{\ell,j}$).
The electron hopping energy from site ($\ell', j'$) to site
($\ell, j$) is described by $t_{(\ell,j),(\ell',j')}$. For AGNRs,
the tight-binding parameters used are $E_{\ell,j} = 0$ for the
on-site energy and $t_{(\ell,j),(\ell',j')} = t_{pp\pi} = 2.7$ eV
for the nearest neighbor hopping strength.

To analyze the electron ballistic transport behavior of AGNR
heterostructures, we need to calculate their transmission
coefficient ${\cal T}_{LR}(\varepsilon)$. This coefficient
describes the probability for each electron trajectory between the
left ($L$) and right ($R$) electrodes. Obtaining the closed-form
expression of ${\cal T}_{LR}(\varepsilon)$ is challenging, even
though $H_{\text{AGNR}}$ does not include electron Coulomb
interactions. These interactions will be accounted for by an
effective two-site Hubbard model later to study charge transport
through the SCTSs. To calculate ${\cal T}_{LR}(\varepsilon)$, we
use the numerical code, which is given by ${\cal
T}_{LR}(\varepsilon) = 4
\text{Tr}[\Gamma_L(\varepsilon)G^r(\varepsilon)\Gamma_R(\varepsilon)G^a(\varepsilon)]$.
Here, $\Gamma_L(\varepsilon)$ and $\Gamma_R(\varepsilon)$
represent the tunneling rates (in energy units) at the left and
right leads, respectively. For simplicity, we consider the wide
band limit for the electrodes and an energy-independent
approximation for $\Gamma_L(\varepsilon)$ and
$\Gamma_R(\varepsilon)$. The retarded and advanced Green functions
of the AGNR, $G^r(\varepsilon)$ and $G^a(\varepsilon)$,
respectively, can be calculated using the system Hamiltonian ($H =
H_0 + H_{\text{AGNR}}$) [\onlinecite{Kuo2},\onlinecite{Kuo3}].

\section{Results and discussion}
\subsection{9-7 AGNR Superlattices}
Topological states (TSs) hold significant promise for applications
in electronics and optoelectronics due to their robust transport
characteristics, which are resistant to defect scattering
[\onlinecite{CohenML}--\onlinecite{ZhaoFZ}]. Recently, there has
been extensive research focused on exploring the topological
phases of various graphene nanoribbons
(GNRs)[\onlinecite{CohenML}--\onlinecite{ZhaoFZ}]. The first
principle method, also known as density functional theory (DFT),
has been employed to calculate the Zak phase of 9-7 AGNR
heterojunctions. This calculation confirms the presence of a
localized interface state with energy at the
midgap[\onlinecite{CaoT}]. Notably, the distinctive feature of the
topological phases in the 9-7 AGNR superlattice is the formation
of minibands arising from the interface-localized states, a
phenomenon well-captured by the Su-Schrieffer-Heeger (SSH) model
[\onlinecite{DJRizzo},\onlinecite{SuWP}]. Using the Hamiltonian
$H_{AGNR}$ (Equation (1)), we investigate the electronic
structures of various $9_w-7_x$ AGNR superlattices (SLs) with
different AGNR segments, as illustrated in Fig. 2 (a)-(f). Here,
the subscripts $w$, and $x$ represent the length of each AGNR
segment in terms of unit cells (u.c.). The presence of topological
states at the interfaces between the 9 AGNR segment and 7 AGNR
segment results in the formation of minibands within the energy
gap between the conduction and valence subbands. These minibands
can be effectively described using the Su-Schrieffer-Heeger (SSH)
model [\onlinecite{DJRizzo},\onlinecite{SuWP}], which provides a
analytical expression of $E_{SSH}(k)= \pm
\sqrt{t^2_x+t^2_w-2t_xt_w~cos(k\pi/L)}$. In this expression, $t_x$
and $t_w$ represent the electron hopping strengths in the 7 AGNR
segment and 9 AGNR segment, respectively, while $L$ denotes the
length of the super unit cell. The results depicted in Fig. 2
highlight the tunability of $t_x$ and $t_w$ in $9_w-7_x$ AGNR SLs
, making them an excellent platform for investigating the phases
of the SSH model, which could potentially host Majorana Fermions
and superconducting states[\onlinecite{YuXL}].

\begin{figure}[h]
\centering
\includegraphics[angle=0,scale=0.3]{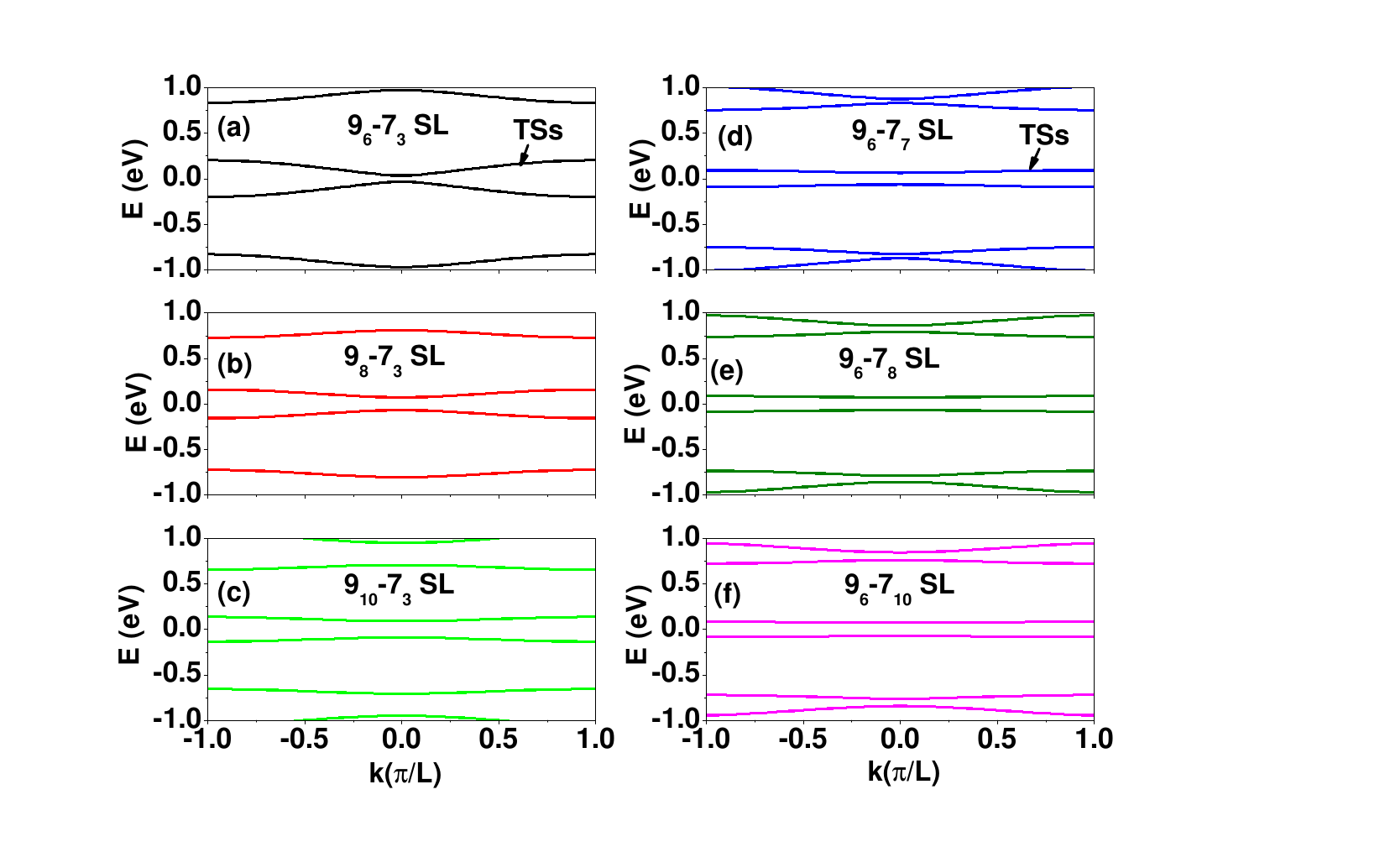}
\caption{Electronic structures of $9_w - 7_x$ AGNR superlattices
(SLs) within the energy range of $|E| \le 1$ eV. (a) $9_6 - 7_3$,
(b) $9_8 - 7_3$, (c) $9_{10} - 7_3$, (d) $9_{6} - 7_7$, (e) $9_{6}
- 7_8$, and (f) $9_{6} - 7_9$ AGNR SLs.}
\end{figure}

\subsection{Su-Schrieffer-Heeger Model}

We have replicated the electronic structures of the minibands
depicted in Fig. 2(a)-2(f) and represented them in Fig. 3(a)-3(f)
using the $E_{SSH}(k)$ expression. Each diagram in Fig. 3
corresponds to its respective counterpart in Fig. 2. The curves in
Fig. 3 are computed using $E_{SSH}(k)$ for specific parameter
values: (a) $t_w = 0.084$ eV, (b) $t_w = 0.0396$ eV, and (c) $t_w
= 0.0176$ eV, with a fixed $t_x = 0.118$ eV. In cases where $w >
x$ (a, b, c), we observe $t_x > t_w$, whereas in cases (d, e, f)
with $w < x$, we find $t_w
> t_x$. The minibands formed in (d), (e), and (f) have narrow
bandwidths, and the corresponding $t_w$ and $t_x$ values in
$E_{SSH}(k)$ are as follows: (d) $t_x = 13.77$ meV, (e) $t_x = 8$
meV, and (f) $t_x = 4.69$ meV, all with a fixed $t_w = 78$ meV.
The band gaps created by these minibands, as shown in Fig.
3(a)-3(f), are as follows: (a) $\Delta_{TS} = 68$ meV, (b)
$\Delta_{TS} = 157$ meV, (c) $\Delta_{TS} = 202$ meV, (d)
$\Delta_{TS} = 129$ meV, (e) $\Delta_{TS} = 140$ meV, and (f)
$\Delta_{TS} = 147$ meV, respectively. Notably, the band gaps
formed by the minibands in 9-7 AGNR superlattices are much smaller
than those created by the topological states of other types of
GNRs [\onlinecite{Groning},\onlinecite{SunQ}]. It is worth noting
that the values of $t_x$ and $t_w$ in 9-7 AGNR superlattices, as
determined using the tight-binding method, might not be as
accurate as those obtained through the DFT calculations
[\onlinecite{DJRizzo}]. Nevertheless, the tight-binding method
still captures the essential characteristics of 9-7 AGNR
superlattices. The examples illustrated in Figures 2 and 3 serve
to demonstrate these outcomes.

\begin{figure}[h]
\centering
\includegraphics[angle=0,scale=0.3]{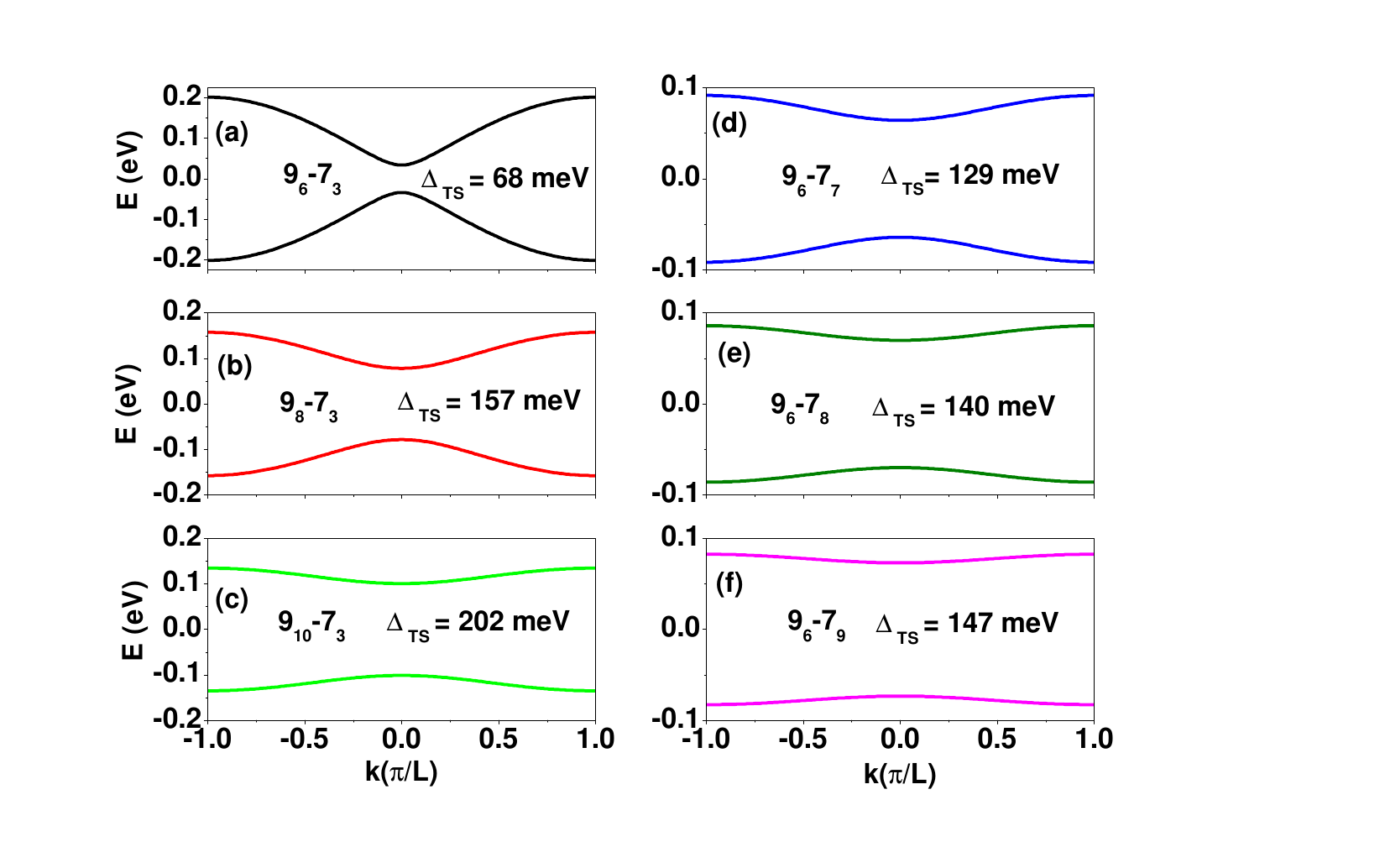}
\caption{Electronic structures of Su-Schrieffer-Heeger model.(a)
$t_w = 0.084$~eV, (b) $t_w = 0.0396$~eV, and (c) $t_w = 0.017$~eV
at a fixed $t_x = 0.118$~eV. (d) $t_x = 13.67$ meV, (e) $t_x =
8$~meV, and (f) $t_x = 4.725$~meV at a fixed $t_w = 78$~meV.}
\end{figure}

The charge transport through a finite 1D SSH chain coupled to
electrodes can offer valuable insights into the charge transport
through the topological states of 9-7-9 AGNR heterostructures. To
investigate this, we employ the finite-chain SSH model and present
the calculated transmission coefficient ${\cal
T}_{LR}(\varepsilon)$ in Fig. 4 at
$\Gamma_{L}=\Gamma_{R}=\Gamma_{t}=10$~meV. The ${\cal
T}_{LR}(\varepsilon)$ curves of Fig. 4(a)-4(c)correspond to the
cases of Fig. 3(a)-3(c), respectively. These ${\cal
T}_{LR}(\varepsilon)$ spectra reflect their band widths and band
gaps in Fig. 3(a)-3(c). Interestingly, the areas of these ${\cal
T}_{LR}(\varepsilon)$ curves exhibit arch-like shapes, rather than
rectangular shapes, indicating that electronic states near the
band edges of minibands have a low probability of transport
between the electrodes. This behavior can be understood by the
electron group velocities at the band edges.

In Fig. 4(a)-4(c), we investigate a finite SSH model with 64 unit
cells. Subsequently, we analyze the impact of finite size on the
SSH model's ${\cal T}_{LR}(\varepsilon)$ due to the limited length
of the 9-7 AGNR heterostructures synthesized using the bottom-up
technique. The heterostructures $9_6-7_7-9_6$ display transport
channels (Fig. 4(d)) resulting from the serially coupled
topological states (SCTS) corresponding to $E_{SSH}(t_w = 0)$,
which are solely determined by the energies of $\varepsilon_{HO}$
and $\varepsilon_{LU}$. Moreover, Fig. 4(e) illustrates ${\cal
T}_{LR}(\varepsilon)$ for $9_6-7_7-9_6-7_7-9_6$ AGNR
heterostructures, presenting four transport channels. Among these,
two channels originate from the outer topological states with an
effective weak hopping strength $t_{eff,x}$, while the other two
channels arise from the topological states of $7-9-7$ with $t_w =
78$~meV [\onlinecite{DJRizzo}] and effective tunneling rates
$\Gamma_{e,t} \ll \Gamma_{t}$. Notably, the probability of
transport through $\Sigma_{outer-TSs}$ is significantly suppressed
due to $\Gamma_{t} > t_{eff,x} = 2.1$~meV. The effective coupling
strengths ($t_{eff,x}$) between the outer TSs of AGNR
heterostructures have been observed in [\onlinecite{Kuo5}]. If one
tunes $\Gamma_{t}$ to $1$~meV, $\Sigma_{outer,TSs}$ will exhibit
two peaks with a separated energy of $|2t_{eff,x}|$. This suggests
that the electron transport between the electrodes is considerably
influenced by the coupling strengths between the AGNR
heterostructures and the electrodes, which, in turn, are
determined by the contact types of electrodes [\onlinecite{Kuo2},
\onlinecite{Kuo3}].

When considering AGNR heterostructures with three 7-AGNR segments
in Fig. 4(f), charge transport through the outer topological
states is almost completely blocked as $\Gamma_{t}\gg
t_{eff,x}=0.399$~meV. In this scenario, the four dominant
transport channels emerge around $\pm t_w$. In Fig. 4(e) and 4(f),
there are $21$ u.c. and $44$ u.c. between the outer TSs (or outer
9 AGNR segments), respectively, and their $t_{eff,x}$ values are
much larger than the $t_x$ values of $9_6-7_{21}-9_6$ and
$9_6-7_{44}-9_6$ (see Fig. 7(a)). This phenomenon is known as the
long-distance coherent tunneling mechanism
[\onlinecite{BraakmanFR}--\onlinecite{Kuo7}]. When $t_w
> t_x$, the outer topological states of AGNR heterostructures
transform into soliton states, provided the AGNR superlattices
have a sufficient length [\onlinecite{SuWP}].

\begin{figure}[h]
\centering
\includegraphics[angle=0,scale=0.3]{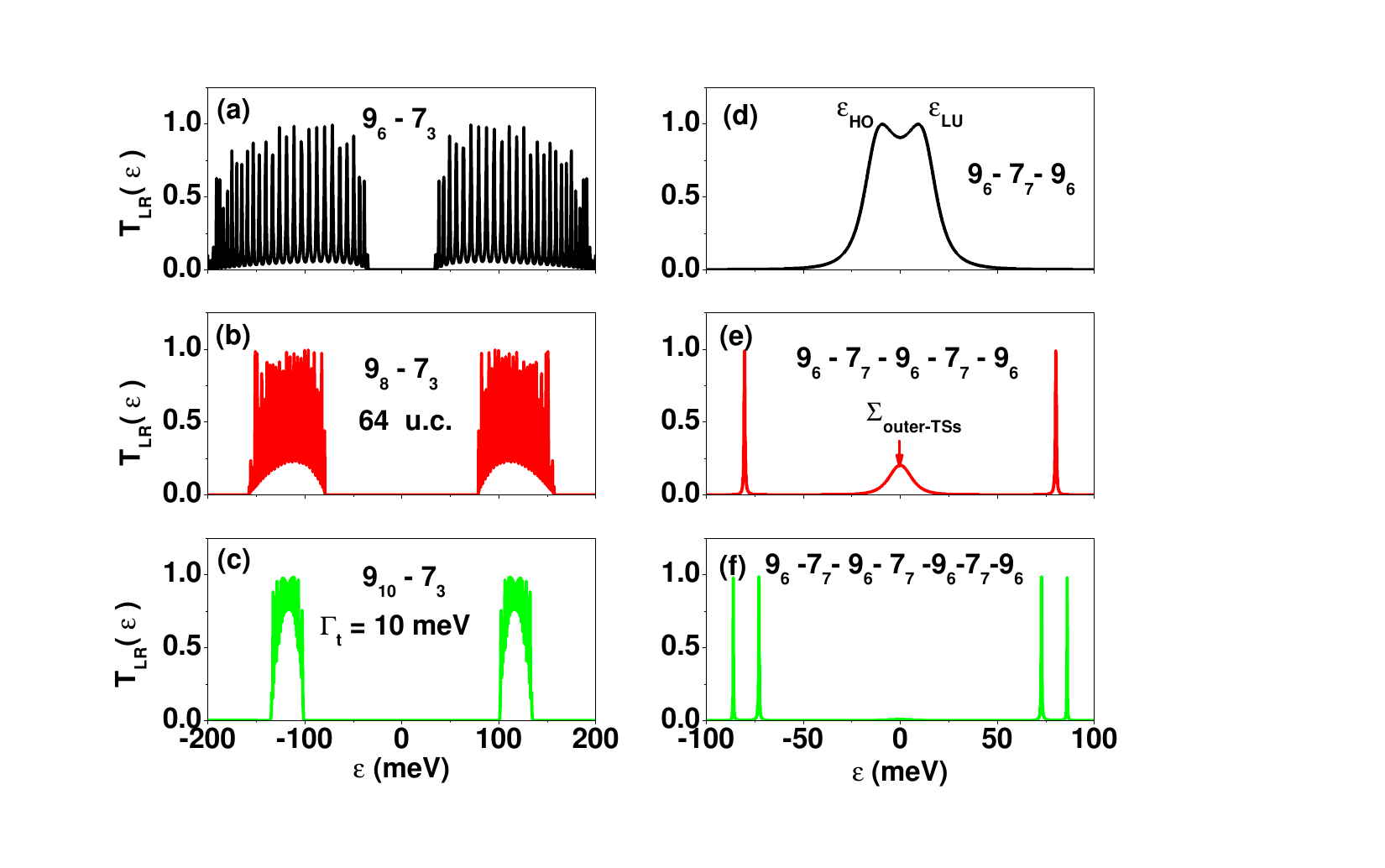}
\caption{Transmission coefficient as a function of $\varepsilon$
for three different scenarios of AGNR superlattices at $\Gamma_{t}
= 10$~meV. (a) $9_6-7_3$, (b) $9_8-7_3$, and (c) $9_{10}-7_3$
superlattices. The values of $t_x$ and $t_w$ are the same as those
in Fig. 3(a)-3(c). Transmission coefficient as a function of
$\varepsilon$ for three different scenarios of AGNR
heterostructures at $\Gamma_{t} = 10$~meV. (d) $9_6-7_7-9_6$, (e)
$9_6-7_7-9_6-7_7-9_6$, and (f) $9_6-7_7-9_6-7_7-9_6-7_7-9_6$ AGNR
heterostructures. The hopping parameters used are $t_x =
13.67$~meV and $t_w = 78$~meV.}
\end{figure}

\subsection{$9_w -7_x-9_y$ AGNR heterostructures }
Although extensive experimental and theoretical efforts have been
dedicated to studying AGNR heterostructures, the effects of
contacts and vacancies on the TSs of asymmetrical  $9_w -7_x-9_y$
AGNR heterostructures have not been clearly
revealed[\onlinecite{ChenYC}--\onlinecite{SunQ}]. In Figure 5, we
present the calculated transmission coefficient ${\cal
T}_{LR}(\varepsilon)$ for $9_{w}-7_x-9_{y}$ AGNR heterostructures
under four different scenarios with
$\Gamma_L=\Gamma_R=\Gamma_t=0.81$ eV. The length of the 7-AGNR
segment remains constant at $x = 3$ u.c., while the left and right
9-AGNR segments have symmetrical and asymmetrical lengths.

In Figure 5(a), a symmetrical situation is depicted, showing two
peaks labeled as $\varepsilon_{HO}$ and $\varepsilon_{LU}$
originating from the TSs with a coupling strength of $t_{x}$
(similar to Fig. 4(d)). The energy separation between
$\varepsilon_{HO}$ and $\varepsilon_{LU}$ is determined by
$|2t_{x}| = 0.2358$ eV, with their maximum values reaching one.
This indicates that the maximum electrical conductance is one
quantum conductance ($G_0$), where $G_0 = \frac{2e^2}{h}$. The
energy levels of $\varepsilon_{HO}$ and $\varepsilon_{LU}$ are
well separated from the bulk states within regions $\varepsilon >
E_c$ and $\varepsilon < E_v$, where $E_c$ and $E_v$ denote the
minimum of the conduction subband and the maximum of the valence
subband (or see Fig. 2(a)-2(c)). The localized charge density of
$\varepsilon_{LU}$ (or $\varepsilon_{HO}$) with $\pm 0.1179$ eV is
plotted in Figure 1(a). From the charge densities of
$\varepsilon_{HO}$ and $\varepsilon_{LU}$, SCTSs function as
serially coupled quantum dots, each containing only one energy
level. Therefore, it is expected that SCTSs can have promising
applications in charge and spin quantum bits
[\onlinecite{DJRizzo},\onlinecite{SunQ},\onlinecite{GurvitzS}--\onlinecite{RuskovR}].

\begin{figure}[h]
\centering
\includegraphics[angle=0,scale=0.3]{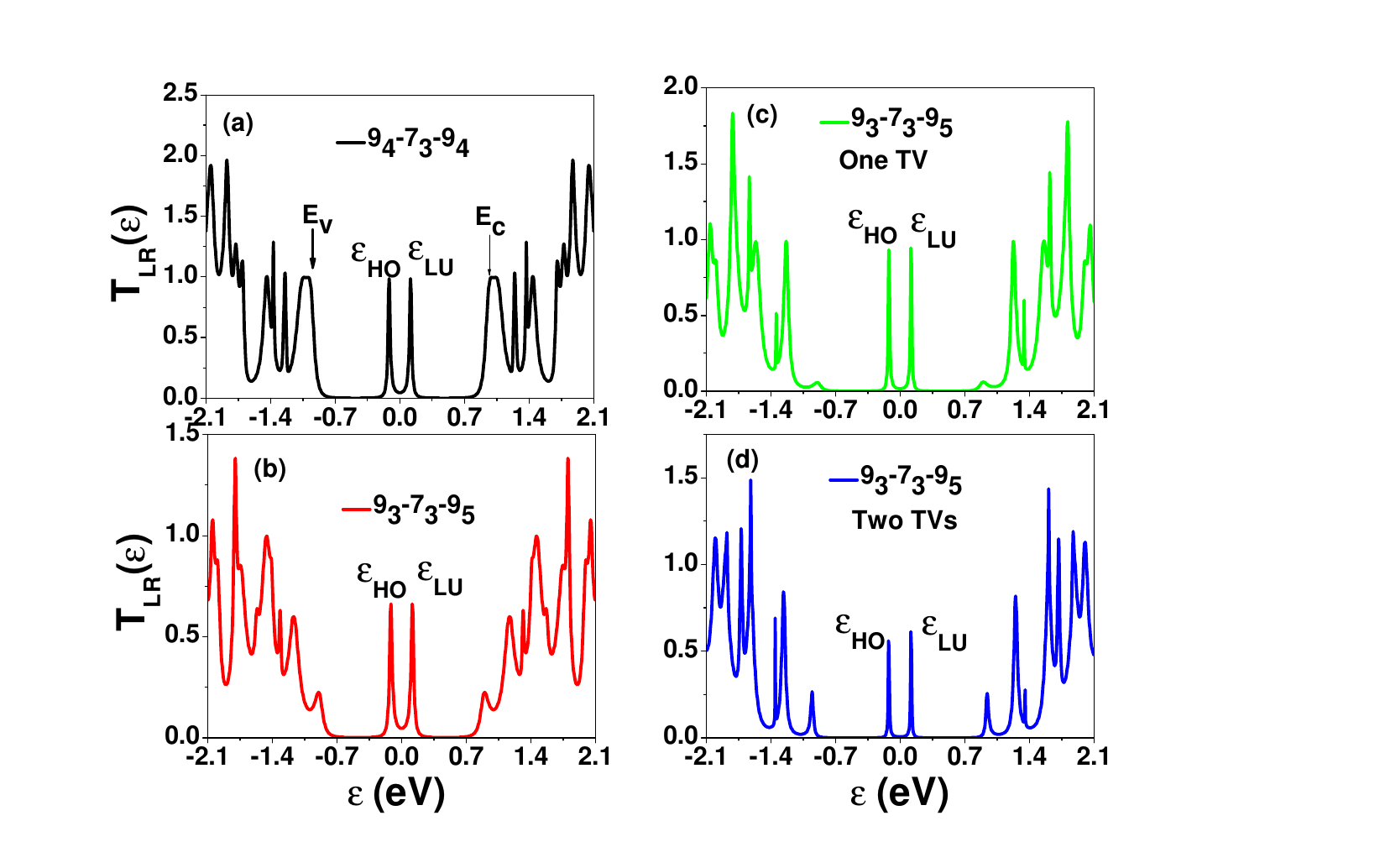}
\caption{Transmission coefficient as functions of $\varepsilon$
for different $9_w-7_x-9_y$ AGNR heterostructures with total
lengths $N_a = 44$ (or $L_a=4.344$~nm) at
$\Gamma_L=\Gamma_R=\Gamma_t=0.81$~eV. (a) $9_4-7_3-9_4$, (b)
$9_3-7_3-9_5$ , (c) $9_3-7_3-9_5$ with one triple vacancy (TV) and
(d) $9_3-7_3-9_5$ with two triple vacancies (TVs). The sites of
vacancies are shown in Fig. 1(b) and 1(c).}
\end{figure}

Figure 5(b) shows that the peak positions of $\varepsilon_{HO}$
and $\varepsilon_{LU}$ remain robust against variations in the
9-AGNR segments, although the strength of these two peaks and the
"bulk states" experience significant changes. The magnitudes of
$\varepsilon_{HO}$ and $\varepsilon_{LU}$ are reduced to
$\frac{2}{3}$. The variation in the 9-AGNR segments serves to
adjust the "barrier width" between the TSs and the electrodes.
This result indicates that the left TS (LTS) and the right TS
(RTS) have different effective coupling strengths to the
electrodes. When a triple vacancy (removing three carbon atoms
shown in Fig. 1(b)) appears between the left electrode and the
LTS, the peaks of $\varepsilon_{HO}$ and $\varepsilon_{LU}$ become
narrower, as observed in Fig. 5(c). The role of the triple vacancy
is to increase the "barrier height" between the left electrode and
the LTS. Consequently, the widths of $\varepsilon_{HO}$ and
$\varepsilon_{LU}$ become narrower, while their energy levels
remain unchanged. In the case of two triple vacancies in
$9_3-7_3-9_5$ AGNR heterostructures, as illustrated in Fig. 1(c),
the $\varepsilon_{HO}$ and $\varepsilon_{LU}$ peaks are further
narrowed, as shown in Fig. 5(d). The results in Fig. 5 demonstrate
the robustness of the TSs against vacancies when the vacancies
occur in regions with low charge densities. These vacancies
effectively suppress phonon transportation and reduce phonon heat
currents [\onlinecite{JWJiang}--\onlinecite{SHTan}].

In Figure 5, the spectra of the transmission coefficient
demonstrate electron-hole symmetry. To enable heat rectification
in practical applications, it is crucial to break the degeneracy
between the LTS and the RTS. i.e., $|\varepsilon_{HO}| \neq
\varepsilon_{LU}$. To achieve this, we introduce a right gate
electrode that modulates the energy levels of the RTS associated
with the $p_z$ orbital. The designed right gate electrode covers
the carbon atoms within the region between $ j = (w + x - 1)
\times 4$ and $j = (w + x + 1) \times 4$. In Fig. 6(a) and 6(b),
we present the calculated transmission coefficient ${\cal
T}_{LR}(\varepsilon)$ for $9_8-7_x-9_6$ and $9_6-7_x-9_8$ AGNR
heterostructures, considering $x = 8$ with $V_{gR} = 45$~mV and $x
= 10$ with $V_{gR} = 27$~mV. Upon applying the right gate
electrode voltage, the peak near the $\varepsilon = 0$ corresponds
to the LTS.

To gain insights into the behavior of the ${\cal
T}_{LR}(\varepsilon)$ spectra in Fig. 6(a) and 6(b), we introduce
a two-site model with the following closed-form expression for
${\cal T}^{2-site}_{LR}(\varepsilon)$:

\begin{small}
\begin{eqnarray}
& &{\cal T}^{2-site}_{LR}(\varepsilon)\\ \nonumber &=&
\frac{4\Gamma_{e,L}t^2_{x}
\Gamma_{e,R}}{|(\varepsilon-E_{e,L}+i\Gamma_{e,L})(\varepsilon-E_{e,R}+i\Gamma_{e,R})-t^2_{x}|^2},
\end{eqnarray}
\end{small}

Here, $\Gamma_{e,L}$ and $\Gamma_{e,R}$ represent the effective
tunneling rates of the LTS with energy level $E_{e,L}$ and the RTS
with energy level $E_{e,R}$, respectively. In Fig. 6(c), we used
$t_{x} = 8$~meV, $E_{e,L} = -1.79$~meV, and $E_{e,R} = 27.4$~meV
for $9_8-7_8-9_6$ AGNR heterostructures with $V_{gR} = 45$~mV, and
$t_{x} = 2.7$~meV, $E_{e,L} = 0$~meV, and $E_{e,R} = 15.75$~meV
for $9_8-7_{10}-9_6$ AGNR heterostructures with $V_{gR} = 27$~mV.
Comparing the results of Fig. 6(a) and 6(c), we observe that Eq.
(2) accurately captures the spectra of TSs with the gate voltage.
From the results of Fig. 6(c) and 6(d), we find that
$\Gamma_{e,L(R)} = 1$~meV and $\Gamma_{e,L(R)} = 3$~meV for the
LTS (RTS) in $9_w-7_x-9_y$ AGNR heterostructures with $w(y) = 8$
and $w(y) = 6$, considering $\Gamma_t = 0.81$~eV. The results from
Figures 5, and 6 indicate that $\Gamma_{e,L(R)}$ is determined by
three factors: (a) the length of the 9-AGNR segment, (b) the
presence of vacancies, and (c) $\Gamma_{L(R)}$. For a given
$\Gamma_{L(R)}$ value, we can easily adjust $\Gamma_{e,L(R)}$ by
tuning the length of the 9-AGNR segment.

\begin{figure}[h]
\centering
\includegraphics[angle=0,scale=0.3]{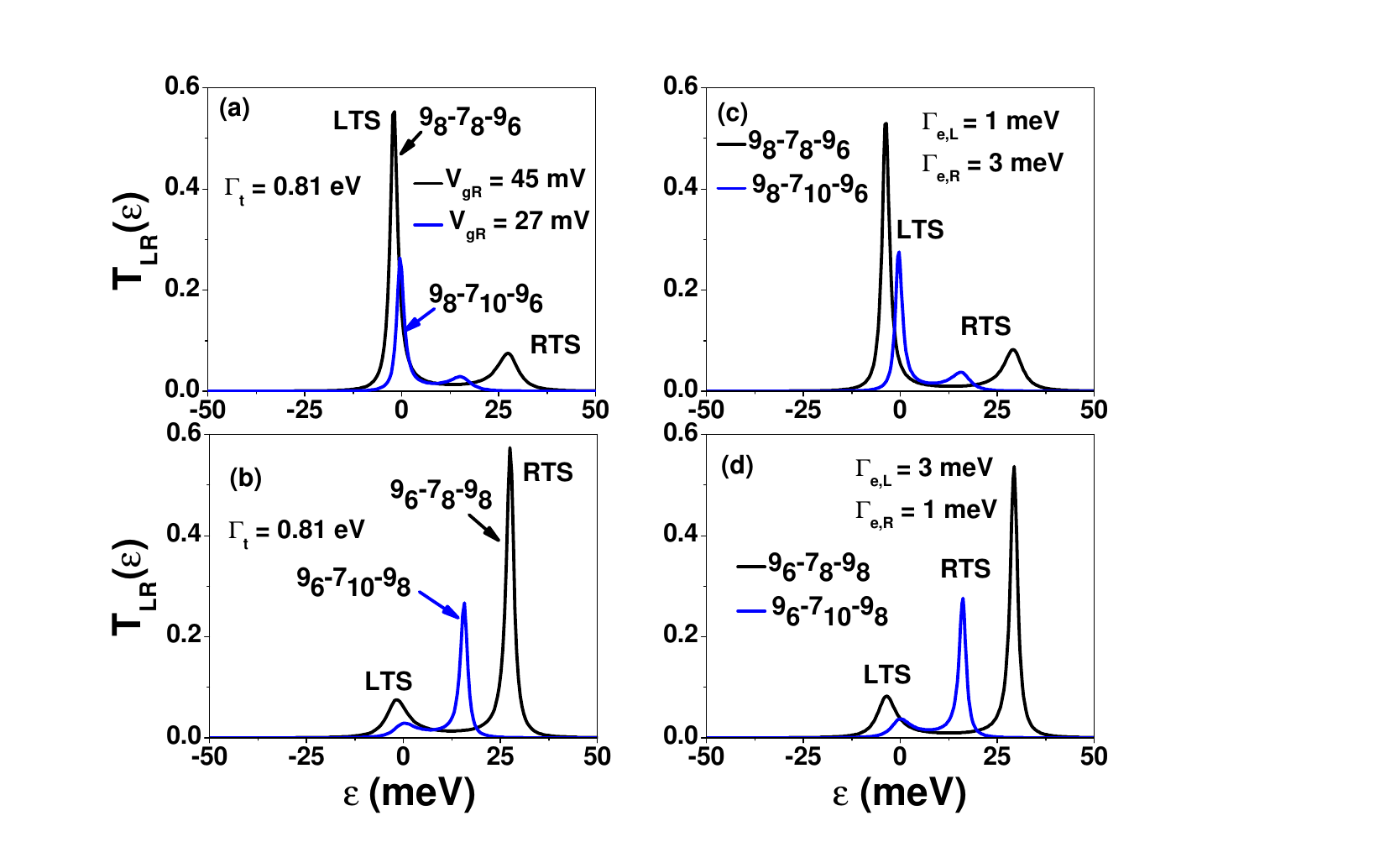}
\caption{Transmission coefficient as functions of $\varepsilon$
for (a) $9_8-7_x-9_6$ AGNR heterostructues and (b) $9_6-7_x-9_8$
AGNR heterostructurs. We have considered $V_{gR} = 45$~mV and
$V_{gR} = 27$~mV for $x = 8$ and $x = 10$, respectively. Addition,
$\Gamma_L = \Gamma_R = \Gamma_t=0.81$~eV in (a) and (b). The
curves of (c) and (d) are calculated by using 2-site model (Eq.
(2)). (c) $\Gamma_{e,L} = 1$~meV and $\Gamma_{e,R} = 3$~meV and
(d) $\Gamma_{e,L} = 3$~meV and $\Gamma_{e,R} = 1$~meV. Other
physical parameters are $t_{x}= 8$~meV, $E_{e,L}=-1.79$~meV, and
$E_{e,R}= 27.4$~meV for $x = 8$. $t_{x}= 2.7$~meV,
$E_{e,L}=0$~meV, and $E_{e,R}= 15.75$~meV for $x = 10$.}
\end{figure}

\subsection{2-site Hubbard model}
Due to the localized wave functions of the TSs, the
electron Coulomb interactions exhibit significant strength. To
elucidate the electron heat current of $9_w-7_x-9_y$ AGNR
heterostructures in the Coulomb blockade region, we employ a
two-site Hubbard model. The Hamiltonian of this model is described
as follows:

\begin{small}
\begin{eqnarray}
& &H_{2-site}\\ \nonumber &=&
\sum_{j=L,R,\sigma}E_{j}c^{\dagger}_{j,\sigma}c_{j,\sigma} -
t_{x}~(c^{\dagger}_{R,\sigma}c_{L,\sigma} +
c^{\dagger}_{L,\sigma}c_{R,\sigma})\\ \nonumber & + &
\sum_{j=L,R}U_j~n_{j,\sigma}n_{j,-\sigma} +
\frac{1}{2}\sum_{j\neq\ell,\sigma,\sigma'}U_{j,\ell}~n_{j,\sigma}n_{\ell,\sigma'},
\end{eqnarray}
\end{small}

where $E_j$ represents the spin-independent energy level of the
TSs, $U_j = U_{L(R)} = U_0$ and $U_{j,\ell} = U_{LR(RL)} = U_1$
denote the intra-TS and inter-TS Coulomb interactions,
respectively, and $n_{j,\sigma} =
c^{\dagger}_{j,\sigma}c_{j,\sigma}$. The values of $U_0$ and $U_1$
are calculated using
$\frac{1}{4\pi\epsilon_0}\sum_{i,j}|\Psi_{L(R)}(\textbf{r}_i)|^2|\Psi_{L(R)}(\textbf{r}_j)|^2\frac{1}{|\textbf{r}_i-\textbf{r}_j|}$
with the dielectric constant $\epsilon_0 = 4$, and $U_{cc} = 4$ eV
at $i = j$. $U_{cc}$ arises from the two-electron occupation in
each $p_z$ orbital of Eq. (1). Here, $\Psi_{L(R)}(\textbf{r}_i)$
represents the wave functions of TSs [\onlinecite{Golor}]. In
Figure 7, we present the calculated electron hopping strengths
between the TSs, as well as the intra-TS and inter-TS electron
Coulomb interactions as functions of $x$ for the $9_3-7_x-9_5$
AGNR heterostructures.

\begin{figure}[h]
\centering
\includegraphics[angle=0,scale=0.3]{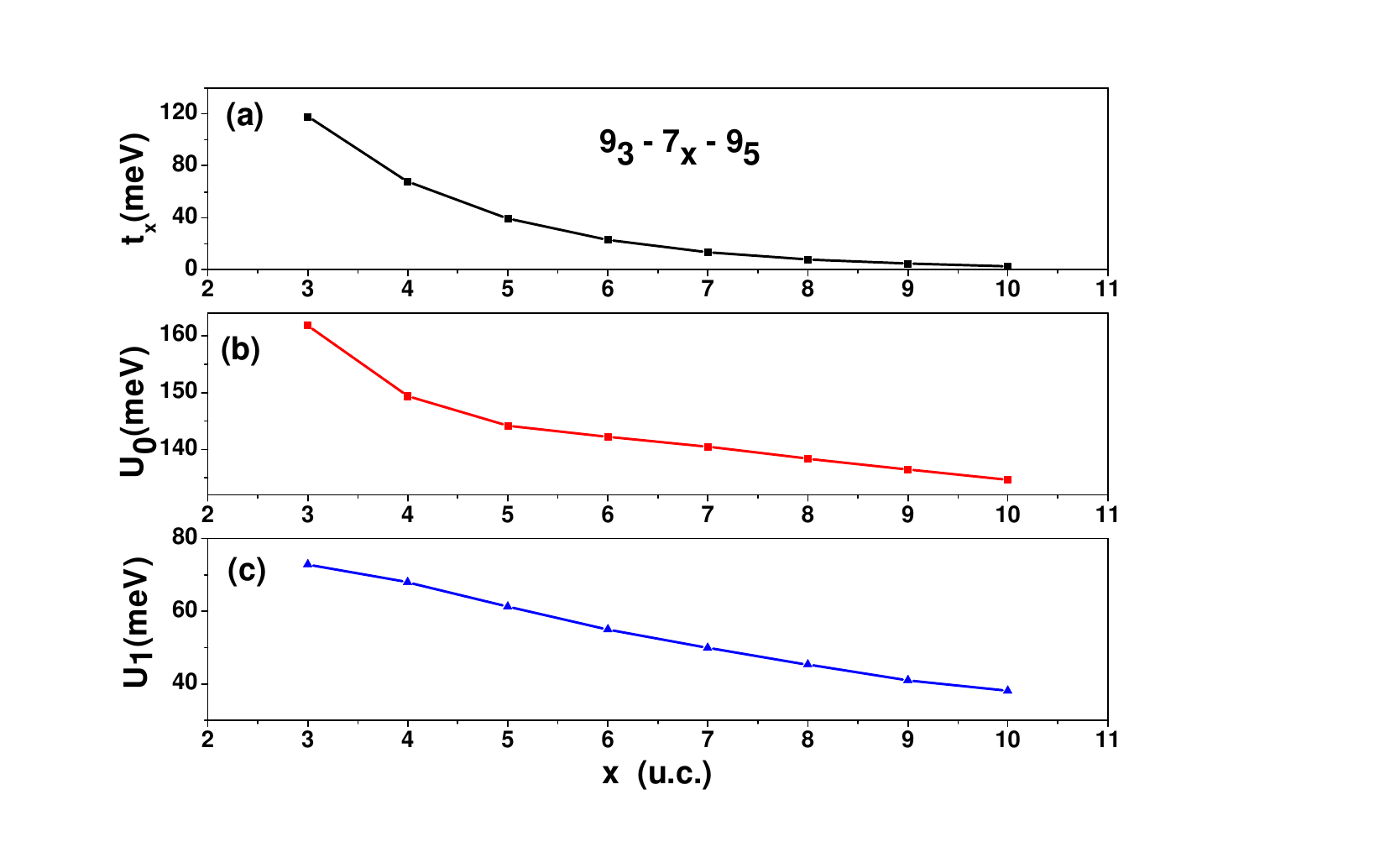}
\caption{(a) Electron hopping strength ($t_{x}$) between the left
TS and the right TS, (b) intra-TS electron Coulomb interactions
($U_0$), and (c) inter-TS electron Coulomb interactions ($U_1$) as
functions of $x$ for $9_3-7_x-9_5$ AGNR heterostructures.}
\end{figure}

Figure 7(a) illustrates that the electron hopping strength
($t_{x}$) decreases rapidly with increasing $x$ since it primarily
depends on the overlap between the localized wave functions of the
left and right TSs ($\Psi_{L}$ and $\Psi_{R}$). On the other hand,
although $U_0$ decreases as $x$ increases, this decrease can be
attributed to the overall increase in the total area of the
$9-7-9$ AGNR. As a result, the sensitivity of $U_0$ to variations
in $x$ is small, particularly when $x > 6$ (Figure 7(b)). This
observation suggests that the charge densities of the TSs are
predominantly distributed at the interfaces (Fig. 1(a)).
Furthermore, Figure 7(c) demonstrates a more rapid decrease in the
inter-TS Coulomb interaction ($U_1$) compared to $U_0$ with
increasing $x$. The trend of $U_0 > U_1 > t_{x}$ persists when $x
\geq 3$. Therefore, it is essential to consider the effects of
electron Coulomb interactions on the charge transport through the
TSs of $9_w-7_x-9_y$ AGNR heterostructures.


\subsection{Electron heat current of SCTSs}

The electron and heat currents leaving from the left (right)
electrode are given by
\begin{eqnarray}
&&J_{L(R)}(\Delta T)\\ \nonumber &=&\frac{2e}{h}\int
{d\varepsilon}~ {\cal
T}^{2-site}_{LR(RL)}(\varepsilon)[f_L(\varepsilon)-f_R(\varepsilon)].
\end{eqnarray}
and
\begin{eqnarray}
&&Q_{e,L(R)}(\Delta T)\\ \nonumber &=&\frac{2}{h}\int
{d\varepsilon}~ {\cal
T}^{2-site}_{LR(RL)}(\varepsilon)(\varepsilon-\mu_{L(R)})[f_L(\varepsilon)-f_R(\varepsilon)].
\end{eqnarray}

where the Fermi distribution function of the $\alpha$ electrode is
given by $f_{\alpha}(\varepsilon) = \frac{
1}{\exp\left(\frac{\varepsilon-\mu_{\alpha}}{k_BT_{\alpha}}\right)+1}$.
To discuss electron heat rectification, we consider the
open-circuit condition ($J_{L(R)} = 0$) under a temperature bias
$\Delta T = T_L - T_R$, where $T_L = T_0 + \frac{\Delta T}{2}$ and
$T_R = T_0 - \frac{\Delta T}{2}$. Here, $T_0$ denotes the average
temperature of the junction system. Due to the Seebeck effect (for
finite $\Delta T$), the thermal voltage $V_{th}$ induced by
$\Delta T$ will balance the electrons diffusing from the hot
electrode to the cold electrode, establishing the condition of
open circuit ($J_{L(R)} = 0$). Additionally, the chemical
potentials also depend on the thermal voltage ($\mu_L = \mu +
\frac{eV_{th}}{2}$ and $\mu_R = \mu - \frac{eV_{th}}{2}$, where
$\mu$ is the equilibrium chemical potential of the electrodes).
The transmission coefficient for charge transport through the
SCTSs with $U_0$ and $U_1$, denoted by ${\cal
T}^{2-site}_{LR}(\varepsilon)$, has a closed-form expression given
by

\begin{small}
\begin{eqnarray}
& &{\cal T}^{2-site}_{LR}(\epsilon)/(4t^2_{x}\Gamma_{e,L}\Gamma_{e,R})=\frac{P_{1} }{|\epsilon_L\epsilon_R-t^2_{x}|^2} \nonumber \\
&+& \frac{P_{2} }{|(\epsilon_L-U_{LR})(\epsilon_R-U_R)-t^2_{x}|^2} \nonumber \\
&+& \frac{P_{3} }{|(\epsilon_L-U_{LR})(\epsilon_R-U_{LR})-t^2_{x}|^2} \label{TF} \\
\nonumber &+&
\frac{P_{4} }{|(\epsilon_L-2U_{LR})(\epsilon_R-U_{LR}-U_R)-t^2_{x}|^2}\\
\nonumber &+& \frac{P_{5} }{|(\epsilon_L-U_{L})(\epsilon_R-U_{LR})-t^2_{x}|^2}\\
\nonumber &+& \frac{P_{6}
}{|(\epsilon_L-U_L-U_{LR})(\epsilon_R-U_R-U_{LR})-t^2_{x}|^2}\\
\nonumber &+&
\frac{P_{7} }{|(\epsilon_L-U_L-U_{LR})(\epsilon_R-2U_{LR})-t^2_{x}|^2}\\
\nonumber
 &+&
\frac{P_{8} }{|(\epsilon_L-U_L-2U_{LR})(\epsilon_R-U_R-2U_{LR})-t^2_{x}|^2}, \\
\nonumber
\end{eqnarray}
\end{small}
where
\begin{small}
\begin{eqnarray}
P_{1}&=&1-N_{L,\sigma}-N_{R,\sigma}-N_{R,-\sigma}+ \langle
n_{R,\sigma}n_{L,\sigma}\rangle \nonumber \\ &+&\langle
n_{R,-\sigma}n_{L,\sigma}\rangle+\langle
n_{R,-\sigma}n_{R,\sigma}\rangle-\langle
n_{R,-\sigma}n_{R,\sigma} n_{L,\sigma} \rangle \nonumber \\
P_{2}&=&N_{R,\sigma}-\langle n_{R,\sigma} n_{L,\sigma}\rangle
-\langle n_{R,-\sigma} n_{R,\sigma}\rangle \nonumber \\ &+&\langle
n_{R,-\sigma} n_{R,\sigma} n_{L,\sigma}\rangle \nonumber \\
P_{3}&=&N_{R,-\sigma}-\langle n_{R,-\sigma} n_{L,\sigma}\rangle
-\langle n_{R,-\sigma}n_{R,\sigma}\rangle \nonumber \\ &+&\langle
n_{R,-\sigma}n_{R,\sigma} n_{L,\sigma} \rangle \nonumber \\
P_{4}&=&\langle n_{R,-\sigma}n_{R,\sigma}\rangle-\langle
n_{R,-\sigma}n_{R,\sigma} n_{L,\sigma}\rangle \nonumber\\
P_{5}&=&N_{L,\sigma}- \langle n_{R,\sigma}n_{L,\sigma}\rangle
-\langle n_{R,-\sigma} n_{L,\sigma}\rangle \nonumber \\ &+&\langle
n_{R,-\sigma}n_{R,\sigma} n_{L,\sigma}\rangle \nonumber \\
P_{6}&=&\langle n_{R,\sigma} n_{L,\sigma}\rangle -\langle
n_{R,-\sigma}n_{R,\sigma} n_{L,\sigma}\rangle \nonumber \\
P_{7}&=&\langle n_{R,-\sigma}n_{L,\sigma}\rangle -\langle
n_{R,-\sigma}n_{R,\sigma} n_{L,\sigma}\rangle \nonumber \\
p_{8}&=&\langle n_{R,-\sigma}n_{R,\sigma}n_{L,\sigma} \rangle
\nonumber,
\end{eqnarray}
\end{small}
where we have $\epsilon_L = \varepsilon-E_{e,L}+i\Gamma_{e,L}$ and
$\epsilon_R = \varepsilon-E_{e,R}+i\Gamma_{e,R}$ in Eq. (6). The
intra-TS and inter-TS two-particle correlation functions are
denoted by $\langle n_{\ell,-\sigma}n_{\ell,\sigma}\rangle$ and
$\langle n_{\ell,\sigma}n_{j,\sigma}\rangle$ ($\langle
n_{\ell,-\sigma}n_{j,\sigma}\rangle$), respectively. $\langle
n_{\ell,-\sigma}n_{\ell,\sigma}n_{j,\sigma}\rangle$ is the
three-particle correlation function. These correlation functions
can be solved self-consistently[\onlinecite{Kuo5}]. While DFT
serves as a potent tool for elucidating the electronic structures
of materials, it remains a challenge to accurately portray
electron transport within the Coulomb blockade region
[\onlinecite{CohenML}--\onlinecite{ZhaoFZ}]. The intricate
interplay of many-body effects gives rise to two-particle and
three-particle correlation functions, posing formulation
difficulties within the mean-field theory framework
[\onlinecite{Joost}].

The closed-form expression of Equation (6) presents a
comprehensive explanation for the transport properties of serially
coupled localized states in both the Coulomb blockade and PSB
regions [\onlinecite{FranssonJ}--\onlinecite{Kondo}]. When there
are no intra-TS and inter-TS Coulomb interactions, equation (6)
simplifies to the 2-site free electron model (equation (2)), which
lacks temperature bias-dependent variables. As a result, the free
electron model fails to exhibit electron heat rectification
behavior, even when an orbital offset exists between the left and
right TS, as depicted in the transmission coefficients of Figure
6. In contrast, the transmission coefficient described by equation
(6) depends on  $\Delta T$ or applied bias $V_{bias}$, which
arises from the electron occupation number and other correlation
functions. Therefore, the presence of electron Coulomb
interactions, along with the orbital offset, significantly
contributes to elucidating the characteristics of electron and
heat current rectification. The electron and heat current
rectification in the PSB is illustrated in the appendix A.

To illustrate the aforementioned crucial statements, we present
the calculated thermal voltage (a, c) and electron heat current
(b, d) as functions of temperature bias ($\Delta T$) for various
average temperature values in Fig. 8, considering two different
AGNR heterostructures ($9_8-7_{10}-9_8$ in (a, b) and
$9_8-7_9-9_8$ in (c, d)). The physical parameters $t_{x}$, $U_0$,
and $U_1$ for $x = 10$ and $x = 9$ are obtained from the data in
Fig. 7. Additionally, we set $E_{e,L} = E_{e,R} = 0$ meV and
$\Gamma_{e,L} = \Gamma_{e,R} = 1$ meV. Moreover, throughout this
article, we assume $\mu = 0$. The thermal voltage ($V_{th}$) is
determined by the condition of zero electron current ($J_{L(R)} =
0$). In the forward temperature bias ($\Delta T > 0$), the
calculated $V_{th}$ is negative, while it is positive in the
backward temperature bias ($\Delta T < 0$). The nonlinearity of
$V_{th}$ at low average temperature ($T_0 = 48$ K) becomes nearly
linear at $T_0 = 72$ K. The electron heat current is a complex
function of $V_{th}$ and $\Delta T$. The forward electron heat
current ($Q_F$) and the backward heat current ($Q_B$) are
calculated from $Q_L$ and $Q_R$, respectively. Comparing the
curves in Fig. 8(d) with those in Fig. 8(b), it can be observed
that the electron heat current is enhanced with increasing
$t_{x}$, while the electron heat currents exhibit symmetrical
behavior in Fig. 8(b) and 8(d), namely $Q_F = |Q_B|$. This
indicates that the electron occupation numbers and other
correlation functions maintain symmetrical behavior with respect
to $\Delta T$.

\begin{figure}[h]
\centering
\includegraphics[angle=0,scale=0.3]{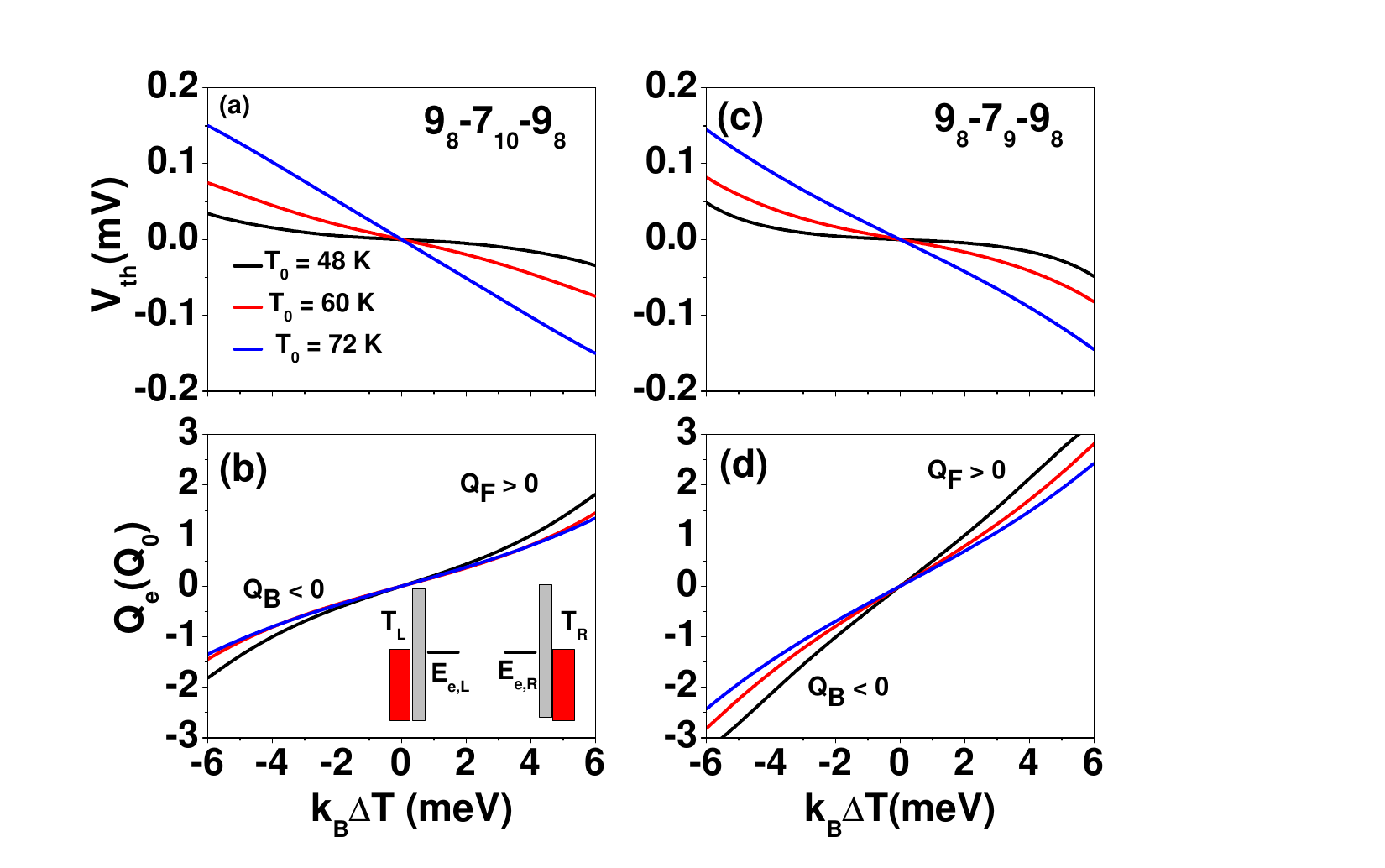}
\caption{ (a) and (b) illustrate the thermal voltage ($V_{th}$)
and electron heat current ($Q_e$) as functions of temperature bias
for various averaged temperature values in $9_8-7_{10}-9_8$ AGNR
heterostructures. (c) and (d) depict the thermal voltage
($V_{th}$) and electron heat current ($Q_e$) as functions of
temperature bias for different averaged temperature values in
$9_8-7_{9}-9_8$ AGNR heterostructures. Here,
$\Gamma_{e,L}=\Gamma_{e,R}=1$~meV and $E_{e,L}=E_{e,R}=0$. The
electron heat current is expressed in units of $Q_0 = 77.2$~pW.}
\end{figure}

To observe asymmetrical electron heat current, it is necessary to
have a temperature bias direction-dependent transmission
coefficient. This implies that $E_{e,L} \neq E_{e,R}$ (or
$\Gamma_{e,L} \neq \Gamma_{e,R}$) is required. Firstly, we
consider the situation where $E_{e,L} \neq E_{e,R}$, which can be
achieved by modulating the gate voltage as shown in Fig. 6. In
Fig. 9, we present the calculated thermal voltage and electron
heat current for various values of $E_{e,R}$, assuming $E_{e,L} =
0$ and $T_0 = 48$ K. As shown in Fig. 9(a), the thermal voltage
$V_{th}$ is significantly enhanced by nearly an order of magnitude
when $E_{e,R} = 10$ meV, compared to the curve in Fig. 8(a) with
$T_0 = 48$ K. However, the thermal voltage remains an odd function
of $\Delta T$, namely $V_{th}(\Delta T) = -V_{th}(-\Delta T)$,
even when $E_{e,L} \neq E_{e,R}$. On the other hand, the electron
heat current $Q_e$ exhibits asymmetrical behavior, where $Q_F \neq
|Q_B|$, as shown in Fig. 9(b). These results indicate that
asymmetrical electron heat current behavior does not necessarily
require asymmetrical thermal voltage behavior.

To highlight the importance of inter-TS Coulomb interaction $U_1$
in the asymmetrical behavior of $Q_e$, we artificially set $U_1$
to zero and plot the calculated $V_{th}$ and $Q_e$ in Fig. 9(c)
and 9(d). It should be noted that $V_{th}$ is less sensitive to
changes in $U_1$, but the asymmetrical behavior of $Q_e$ is
suppressed when $U_1 = 0$. For instance, at $k_B\Delta T = \pm 6$
meV and $E_{e,R} = 10$ meV with $U_1 = 38$ meV, we have $Q_F =
0.934 Q_0$ and $Q_B = -0.384 Q_0$, resulting in a ratio of
$Q_F/|Q_B| = 2.43$. However, when we turn off $U_1$ (i.e., $U_1 =
0$), we obtain $Q_F = 0.936 Q_0$ and $Q_B = -0.738 Q_0$ under the
same conditions, resulting in a ratio of $Q_F/|Q_B| = 1.27$.

\begin{figure}[h]
\centering
\includegraphics[angle=0,scale=0.3]{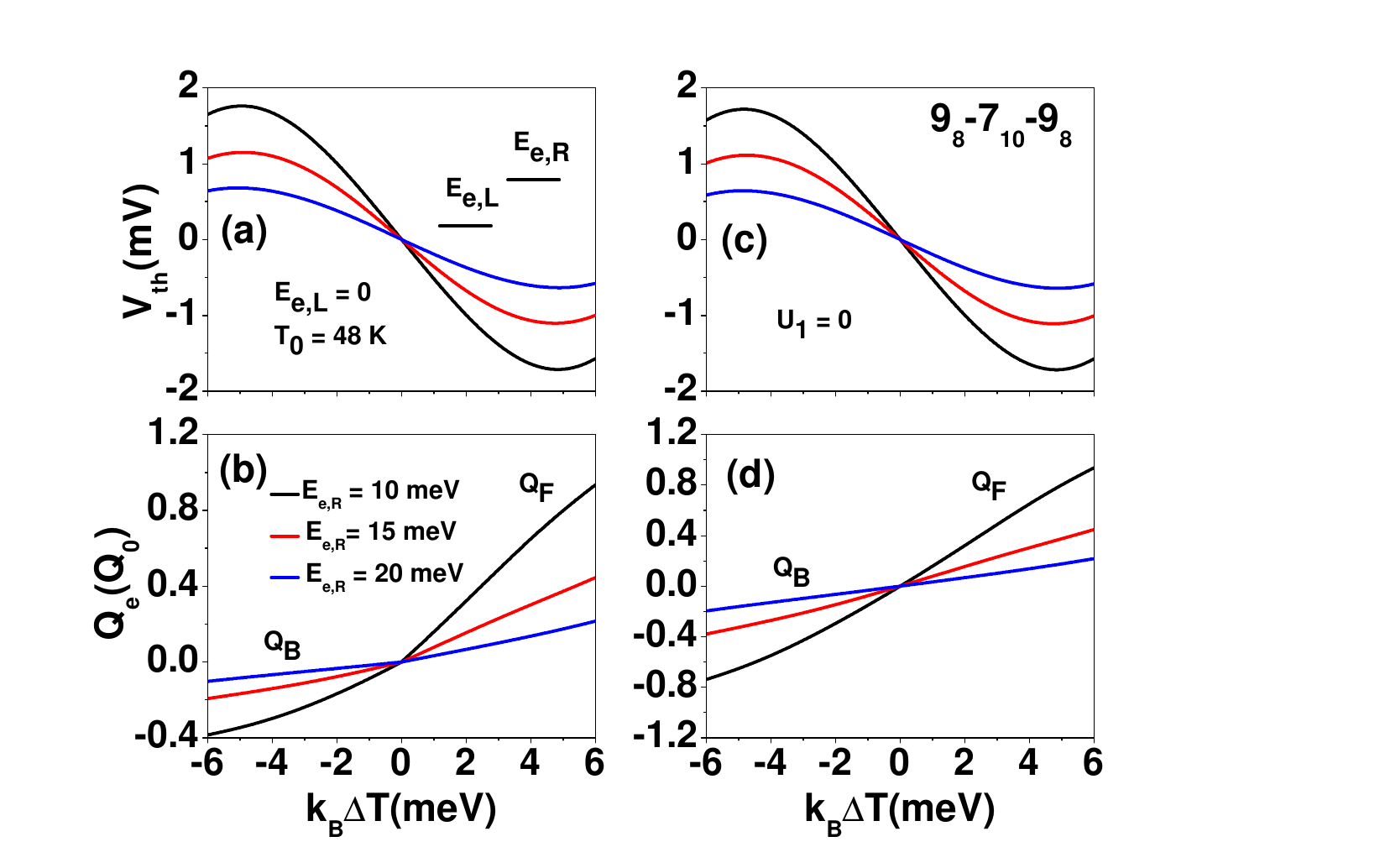}
\caption{(a) and (b) present the thermal voltage ($V_{th}$) and
electron heat current ($Q_e$) as functions of temperature bias for
various $E_{e,R}$ values at $T_0 = 48$~K, with $E_{e,L}=0$, in
$9_8-7_{10}-9_8$ AGNR heterostructures. The curves in (c) and (d),
with $U_1=0$, correspond individually to the curves in (a) and
(b), respectively. Here, $\Gamma_{e,L}=\Gamma_{e,R}=1$~meV. The
electron heat current is measured in units of $Q_0 = 77.2$~pW.}
\end{figure}

Once $E_{e,L} \neq E_{e,R}$, asymmetrical electron heat current
($Q_F > |Q_B|$) can be observed. To investigate the effect of
$\Gamma_{e,L} \neq \Gamma_{e,R}$, we analyze electron occupation
number $N_{j,\sigma}$, electron heat current $Q_e$, and
rectification efficiency $\eta_Q$ for different $\Gamma_{e,L}$
values at $T_0 = 48$ K and $\Gamma_{e,R} = 1$ meV, as shown in
Fig. 10. The variation in $\Gamma_{e,L}$ can be achieved by
adjusting $\Gamma_t$ with different $w$ values and introducing
vacancies. As depicted in Fig. 10(a), $N_{L,\sigma} >
N_{R,\sigma}$ occurs when $E_{e,L}$ is closer to $\mu = 0$.
Additionally, $N_{L,\sigma}$ exhibits significant temperature bias
direction-dependent behavior. These two factors contribute to the
direction-dependent character of the transmission coefficient with
temperature bias $\Delta T$. The asymmetrical electron heat
rectification behavior ($Q_F > |Q_B|$) displayed in Fig. 10(b) can
be understood by considering the $P_1$ channel of Eq. (6).
Notably, the probability weight of $P_1$ in ${\cal
T}_{RL}(\varepsilon)$ under reversed temperature bias can be
obtained by exchanging the indices of $L$ and $R$ in ${\cal
T}_{LR}(\varepsilon)$. Since $E_{e,R}$ is far from $\mu = 0$, the
most influential factors in $P_1$ are the one-particle occupation
numbers, while two-particle and three-particle correlation
functions are negligible. Consequently, the probability weight of
$P_1$ is larger for $\Delta T > 0$ compared to $\Delta T < 0$,
explaining the observed electron heat rectification behavior in
Fig. 10(b).

In Fig. 10(c), the rectification efficiency $\eta_Q$ is enhanced
when considering $\Gamma_{e,L} > \Gamma_{e,R}$. Here, $\eta_Q =
(Q_F+|Q_{ph}|)/(|Q_B|+|Q_{ph}|)-1$. To simplify the calculation,
we calculate $Q_{ph} = F_s \times Q_{\ell} \times \Delta T$, where
$Q_{\ell}$ represents the phonon quantum conductance
($\frac{\pi^2k^2_BT}{3h}$, which is $0.9464 pW/K$ for $T =
1$~K)[\onlinecite{SHTan}], and $F_s = 0.0287$ is the reduction
factor for $Q_{\ell}$ due to the effects of vacancies and
heterostructures [\onlinecite{JWJiang}--\onlinecite{SHTan}].
Notably, $F_s = 0.1198$ provides an excellent fit for the
calculated phonon thermal conductance of AGNR ($\kappa_{ph}$) at
room temperature [\onlinecite{ZhengH}]. The presence of vacancies
and heterostructures in AGNRs not only reduces phonon thermal
conductance but also creates asymmetrical phonon heat currents
[\onlinecite{ZhangZG},\onlinecite{ZhaoH}]. It is worth noting that
electron heat rectification exists even at very small temperature
bias ($\Delta T \rightarrow 0$), indicating that electron thermal
conductances exhibit temperature bias direction-dependent
behavior. This behavior has also been observed in phonon thermal
conductances [\onlinecite{ZhaoH}].

\begin{figure}[h]
\centering
\includegraphics[angle=0,scale=0.3]{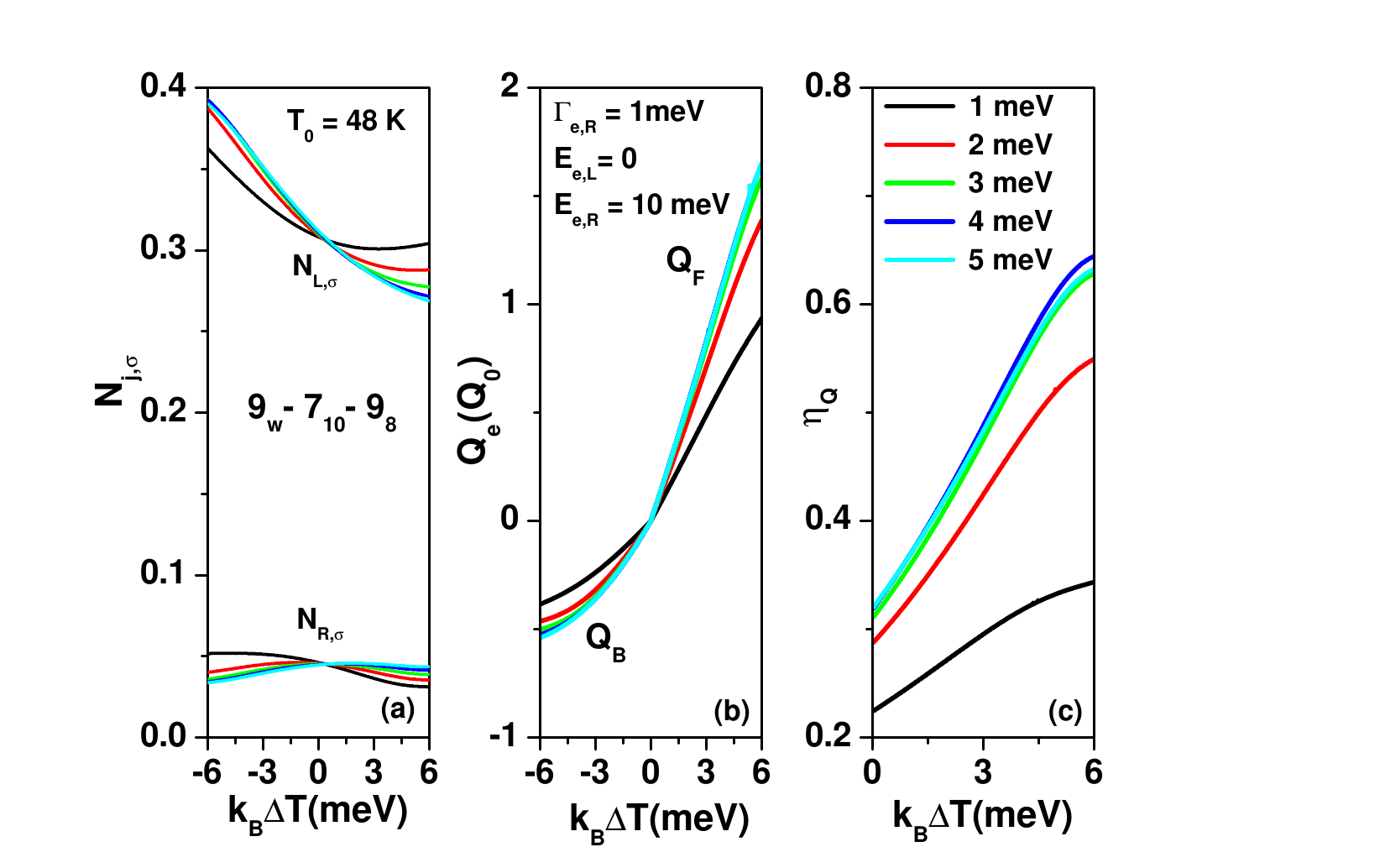}
\caption{(a) Electron occupation number ($N_{j,\sigma}$), (b)
electron heat current ($Q_e$), and (c) rectification ratio
($\eta_Q$) of the $9_w-7_{10}-9_8$ AGNR heterostructure as
functions of temperature bias ($\Delta T$) for various
$\Gamma_{e,L}$ values, with $E_{e,L} = 0$, $E_{e,R} = 10$~meV, and
$T_0 = 48$~K. The heat current is measured in units of $Q_0 =
77.2$~pW.}
\end{figure}

In previous figures (Fig. 9 and Fig. 10) with $t_{x} = 2.7$ meV,
we have demonstrated electron heat current rectification when
$E_{e,L} \neq E_{e,R}$. Now, to investigate the influence of
electron hopping strength between TSs on heat rectification, we
present the thermal voltage $V_{th}$ and electron heat current
$Q_e$ as functions of temperature bias for various 7-AGNR segments
at a low averaged temperature of $T_0 = 24$~K in Fig. 11(a) and
11(b), respectively. As shown in Fig. 11(a), for the case of $x =
8$ with $t_{x} = 8$ meV, $V_{th}$ becomes very small, indicating
that the electron-hole asymmetry lifted by $E_{e,L} \neq E_{e,R}$
weakens for larger $t_{x}$. Moreover, a larger $t_{x}$ results in
a larger electron heat current $Q_e$, as depicted in Fig. 11(b).

To illustrate the rectification efficiency arising from the
enhancement of $t_{x}$, we plotted $\eta_Q$ in Fig. 11(c) and
11(d) with and without the contribution of phonon heat current
$Q_{ph}$. For $T_0 = 24$ K, the rectification efficiency $\eta_Q$
in Fig. 11(c) can reach and exceed 0.5, even for very small
temperature bias. This significant rectification is attributed to
the dominance of electron heat current at low averaged temperature
$T_0$. Furthermore, by artificially turning off $Q_{ph}$ in Fig.
11(d), $\eta_Q$ can reach 1.6 at $k_B\Delta T = 3$ meV for the
case of $x = 10$ with $t_{x} = 2.7$ meV. This result indicates
that $Q_F = 2.6 |Q_B|$ in this case, showcasing a remarkable
rectification effect. Notably, the variation of $t_x$ in Fig. 11
is based on the change in 7 AGNR segment lengths. Additionally,
the long-distance coherent tunneling mechanism
[\onlinecite{BraakmanFR}--\onlinecite{Kuo7}] provides an
alternative means of modifying $t_x$ values. For instance, we can
implement a central gate on the middle 9 AGNR segment of
$9-7-9-7-9$ AGNR heterostructures. By applying a central gate
voltage, we can tune an effective hopping strength $t_{eff,x}$
between the outer topological states of the $9-7-9-7-9$ junctions.
This approach offers further possibilities for controlling and
optimizing the rectification efficiency.

\begin{figure}[h]
\centering
\includegraphics[angle=0,scale=0.3]{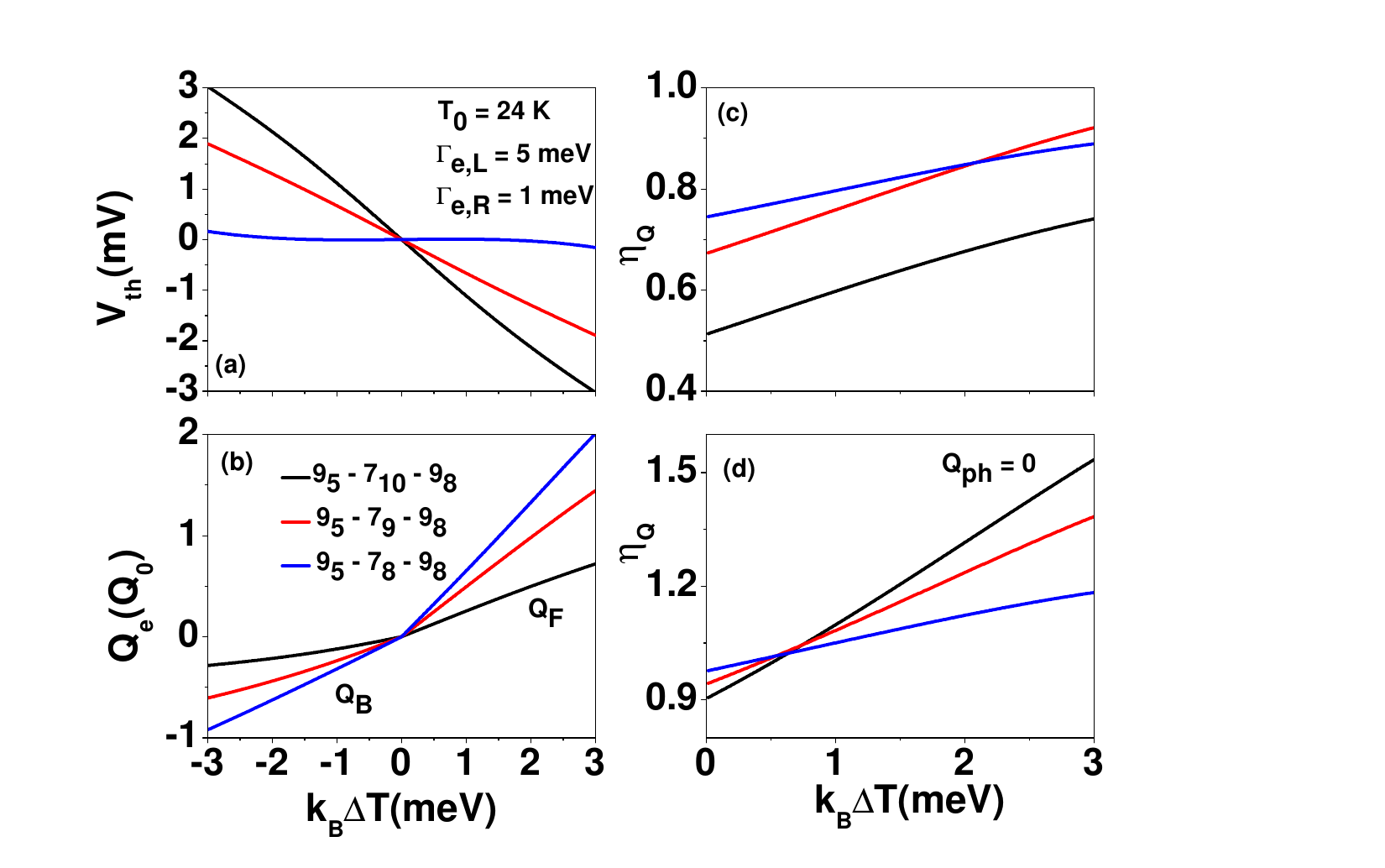}
\caption{(a) illustrates the thermal voltage and (b) depicts the
electron heat current as functions of applied temperature bias for
various $9_5-7_x-9_8$ AGNR heterostructures. The parameters used
are $\Gamma_{e,L}=5$~meV, $\Gamma_{e,R}=1$~meV, $E_{e,L} = 0$,
$E_{e,R}=10$~meV, and $T_0 = 24$~K. (c) and (d) showcase the
electron heat rectification efficiency with and without phonon
heat current ($Q_{ph}$). The heat current is measured in units of
$Q_0 = 77.2$~pW.}
\end{figure}

Based on the findings presented in Fig. 11(c) and 11(d), it
becomes evident that high-efficiency electron heat diodes (HDs)
align with the condition $Q_{e}\gg Q_{ph}$. Consequently, two
strategies emerge: one involves the reduction of $Q_{ph}$ while
preserving $Q_e$, while the other entails enhancing $Q_e$ while
maintaining $Q_{ph}$ at a constant level. In the former case, a
viable approach involves increasing the vacancy density. In the
latter case, a feasible strategy entails utilizing 9-7 AGNR
superlattices in lieu of 9-7-9 AGNR heterostructures. Notably,
earlier work [\onlinecite{HuJN}] highlights that AGNRs containing
one or two vacancies exhibit notably reduced phonon thermal
conductance ($\kappa_{ph}$) when compared to vacancy-free AGNRs.
Particularly, a distinct decline in $\kappa_{ph}$ is observable in
AGNRs harboring two vacancies over the temperature range of 180 K
to 400 K. It is thus reasonable to expect that the rectification
efficiency $\eta_Q$ will be superior for two triple vacancies as
opposed to a solitary triple vacancy, as long as the electron heat
current $Q_e$ remains unaffected by the vacancies. Fundamentally,
the effective realization of proficient electron heat diodes and
thermoelectric devices hinges on the effective management of
phonon heat current. Recent strides in this direction showcase
significant reductions in phonon heat flow within bulk junctions
through the application of strains or magnetic fields
[\onlinecite{ZhangEM},\onlinecite{PanH}]. Considering the context
of diminishing phonon heat flow in 9-7 AGNRs, strain manipulation
proves more advantageous from a device perspective than the
consideration of magnetic fields. Regarding the augmentation of
electron density, it is foreseeable that such an undertaking might
lead to the attenuation of $\eta_Q$ due to the diminished electron
Coulomb interactions within the topological states. Specifically,
the electron Coulomb interactions within the minibands generated
by the topological states, as illustrated in Fig. 2, are
comparatively weaker than those within the $9-7-9$
heterostructures, a consequence attributed to the delocalized
nature of their wave functions.

\section{Conclusion}
In conclusion, our theoretical analysis focuses on achieving
electron heat rectification in charge transport through the
topological states (TSs) of asymmetrical length armchair graphene
nanoribbon (AGNR) heterostructures with vacancies, specifically
the $9_w-7_x-9_y$ configuration.  Our results show that the
topological states (TSs) of 9-7-9 heterostructures are effectively
separated from the subbands, effectively reducing subband noise
caused by temperature fluctuations during electron heat transfer.
The localized charge densities of the TSs at the interfaces
between the $9$-AGNR and $7$-AGNR segments allow for adjusting the
TSs-electrode separation distance by varying the length of the
$9$-AGNR segments, offering control over the coupling strengths
(or effective tunneling rates) between the electrodes and TSs.
Notably, vacancies between the electrodes and TSs do not
compromise the integrity of the TSs but instead aid in suppressing
phonon heat current in the $9_w-7_x-9_y$ AGNR heterostructures.

Moreover, we find that the electron hopping strengths of the TSs
($t_{LR}$) can be tuned by varying the length of the $7$-AGNR
segment or using a long-distance coherent tunneling mechanism. The
localized wave functions of the TSs exhibit significant electron
Coulomb interactions, which decrease gradually as the length of
the $7$-AGNR segment increases, a critical factor for observing
pronounced asymmetrical electron heat currents. This asymmetry
arises from the temperature bias direction-dependent transmission
coefficient resulting from both intra-TS and inter-TS electron
Coulomb interactions, in addition to the asymmetrical TS energy
levels. Furthermore, the optimized orbital offset of the TSs can
be modulated effectively using a designed gate electrode. Our
proposed electron heat diodes achieve a remarkable heat
rectification efficiency ($\eta_Q$) of 0.7 for very small
temperature biases at an averaged temperature of $T_0 = 24$ K in
the $9_5-7_8-9_8$ ($N_a = 84$, $L_a = 8.8$ nm) configuration. This
finding indicates a successful suppression of the short channel
effect.


{}

{\bf Acknowledgments}\\
{This work was supported by the National Science and Technology
Council, Taiwan under Contract No. MOST 107-2112-M-008-023MY2.}

\mbox{}\\
E-mail address: mtkuo@ee.ncu.edu.tw\\

\appendix
\numberwithin{figure}{section}

\numberwithin{equation}{section}

\section{Electron and Heat Currents of $9_8-7_{10}-9_6$ AGNR Heterostructures with PSB Configurations}
\numberwithin{figure}{section}

 \numberwithin{equation}{section}

In this appendix, we investigate the electron and heat currents
associated with charge tunneling through a serially coupled
topological state (SCTS) composed of $9_8-7_{10}-9_6$ AGNR
heterostructures under Pauli spin blockade (PSB) configurations
[\onlinecite{Ono}]. Figure A.1 illustrates the calculated electron
occupation number, two-particle correlation functions, and
tunneling current as a function of applied bias at $T_0 = 12$ K
and $\Delta T = 0$ K. The parameters considered are $E_{e,L}= -U_1
+ \delta_L eV_{\text{bias}}$, $E_{e,R}= -U_0 + \delta_R
eV_{\text{bias}}$, $\Gamma_{e,L}= 1$ meV, and $\Gamma_{e,R}= 3$
meV. In diagrams (a), (b), and (c), $\delta_L = \delta_R = 0$ is
used to ignore the applied bias-dependent orbital offset. In
diagrams (d), (e), and (f), $\delta_L = 1/6$ and $\delta_R = -1/4$
are used, which are determined by the length of each 9-AGNR
segment and the total length of the $9_8-7_{10}-9_6$ AGNR
heterostructure, for simplicity [\onlinecite{Kuo6}]. Additionally,
$E_{e,L}= -U_1$ and $E_{e,R} = -U_0$ are set by introducing two
gate electrodes to modulate the energy levels of the left and
right topological states (TSs). For instance, $V_{g,L} = -81$ mV
and $V_{g,R} = -360$ mV can be set for the $9_8-7_{10}-9_6$ AGNR
heterostructure to obtain $E_{e,L} = -U_1 = -38$ meV and $E_{e,R}
= -U_0 = -134$ meV. It should be noted that the regions covered by
$V_{gL}$ and $V_{gR}$ are the first region from $j = (w-1)*4$ to
$j = (w+1)*4$ and the second region from $j = (w+x-1)*4$ to $j =
(w+x+1)*4$, respectively.

In the case of a forward applied bias ($V_{\text{bias}} > 0$), the
total electron occupation number $N_T =
\sum_{\sigma}(N_{L,\sigma}+N_{R,\sigma})$ approaches two, with
$N_{L,\sigma} = 0.5$ and $N_{R,\sigma} = 0.5$, indicating one
electron occupation in the left and right TSs, respectively (see
Fig. A.1(a)). Conversely, under a backward applied bias, the
electron occupation number $N_{R,\sigma}$ of the right TS exceeds
0.5, while the electron occupation number $N_{L,\sigma}$ of the
left TS is less than 0.5.

Figure A.1(b) displays the curves of the two-particle correlation
functions for inter-TSs with singlet state (2-site-S) and triplet
state (2-site-T). As $V_{\text{bias}} > 0$, the SCTS is
predominantly occupied by triplet states. Conversely, for
$V_{\text{bias}} < 0$, the curves of 2-site-T and 2-site-S merge
together, but they are lifted for $V_{\text{bias}} < -76$ mV
because electrons from the right electrode are tunneling through
the resonant channels resulting from $E_{e,L}+2U_{1} =
E_{e,R}+U_0+U_1$.

In Fig. A.1(c), we observe significant tunneling current
rectification under PSB configurations. The darkness of $J_F$ is
attributed to the SCTS being predominantly occupied by the triplet
state. Furthermore, in Fig. A.1(d), (e), and (f), we reveal the
effect of bias-dependent orbital offset on single-particle
occupation numbers, two-particle correlation functions, and
tunneling current, respectively. The bias-dependent orbital offset
lifts the resonant channels arising from the condition of $E_{e,L}
+ U_1 = E_{e,R} + U_0 = \mu$. Notably, when $E_{e,L} + U_1 +
eV_{\text{bias}}/6 \neq E_{e,R} + U_0 - eV_{\text{bias}}/4 $, a
remarkable effect is observed not only on $N_{j,\sigma}$ and
two-particle correlation functions but also on the tunneling
current in the range of $|V_{\text{bias}}| > 12.5$ mV. The current
rectification spectra shown in Fig. A.1(f) have been
experimentally measured in serially coupled GaAs quantum dots
[\onlinecite{Ono}]. A spin-current conversion device
[\onlinecite{Ono},\onlinecite{TongC}] requires a tunneling current
that can highly resolve the spin singlet and triplet states.
Consequently, it is necessary to suppress the effect of
bias-dependent orbital offset of SCTSs.

\begin{figure}[h]
\centering
\includegraphics[angle=0,scale=0.3]{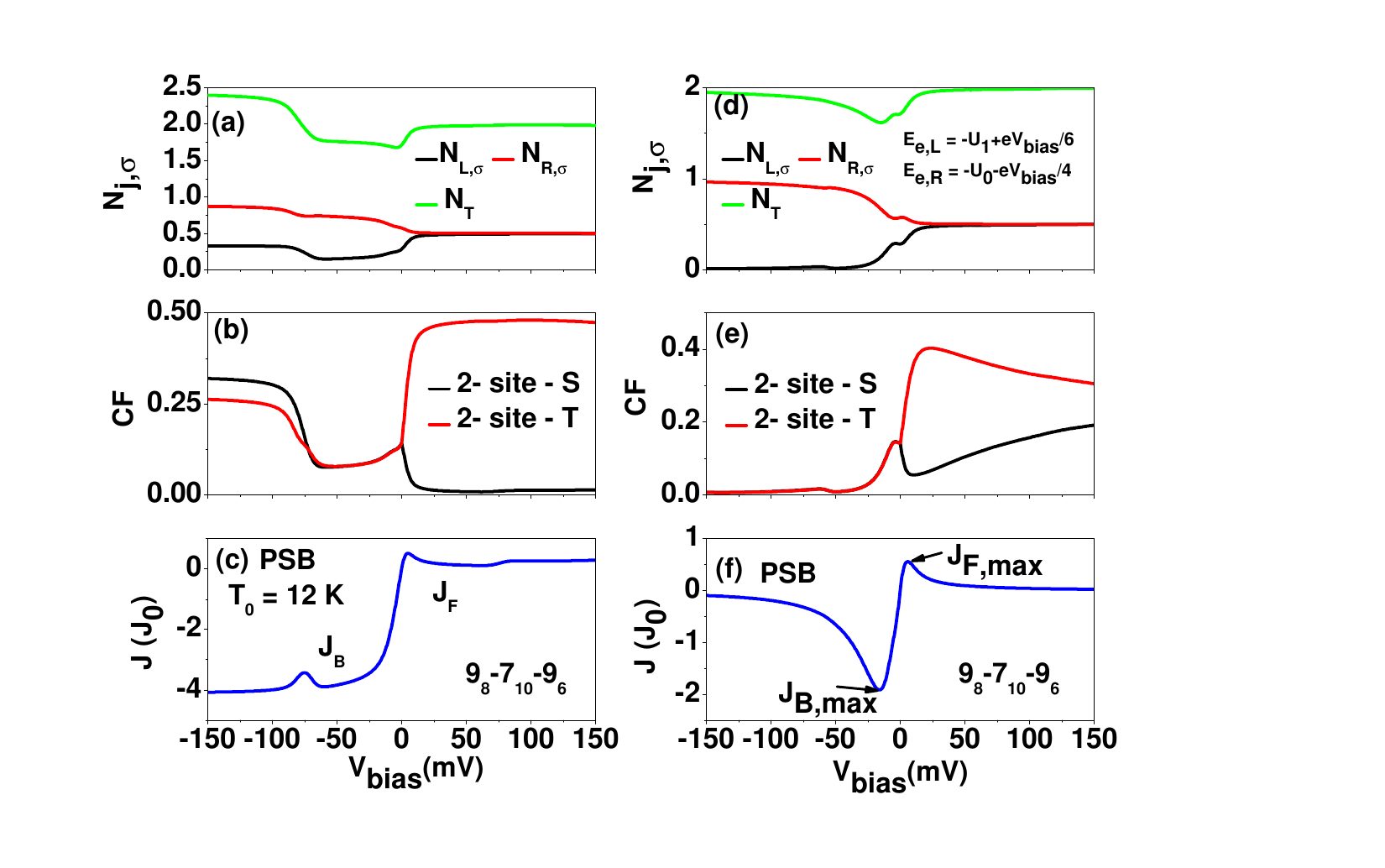}
\caption{(a) Electron occupation number ($N_{j,\sigma}$), (b)
two-particle correlation functions (CF), and (c) tunneling current
($J$) as functions of applied bias $V_{bias}$ at $T_0 = 12$~K and
$\Delta T = 0$~K for $9_8-7_{10}-9_6$ AGNR heterostructures with
Pauli spin blockade (PSB) configurations without considering
bias-dependent orbital offset. In (d),(e) and (f) the effect of
bias-dependent orbital offset is included by considering $E_{e,L}=
-U_1 + eV_{bias}/6$ and $E_{e,R} = -U_0- eV_{bias}/4$. We have
also adopted $U_0 = 134$~meV, $U_1 = 38$~meV and $t_{LR}=2.7$~meV.
The applied bias-dependent left and right chemical potentials are
$\mu_L = \mu + eV_{bias}/2$ and $\mu_R = \mu - eV_{bias}/2$. The
tunneling current is measured in units of $J_0 = 0.773$~nA. }
\end{figure}

Based on the results shown in Fig. A.1, it can be concluded that
the effect of applied bias-dependent orbital offset is significant
only for large applied biases, specifically $|V_{\text{bias}}|>
12.5$ mV. Since the values of $V_{\text{th}}$ are small, we can
safely neglect this effect when discussing the electron heat
current under PSB configurations. To illustrate the electron heat
current in $9_8-7_{10}-9_6$ AGNR heterostructures with PSB
configurations, we need to consider an open circuit condition
($J_{L(R)} = 0$) to determine the thermal voltage
($V_{\text{th}}$) arising from the Seebeck effect and other
self-consistently solved correlation functions. In Fig. A.2, we
present the calculated electron occupation numbers
($N_{j,\sigma}$), thermal voltage ($V_{\text{th}}$), and electron
heat current ($Q_e$) as functions of temperature bias at $T_0 =
24$ K.

As observed in Fig. A.2(a), $N_{R,\sigma} > 0.5$, and
$N_{L,\sigma} < 0.5$. The variation of $N_{j,\sigma}$ with respect
to temperature bias $\Delta T$ is less sensitive. However, the
thermal voltage $V_{\text{th}}$ shown in Fig. A.2(b) is extremely
small compared to the results in Fig. 9. This negligible
$V_{\text{th}}$ can be attributed to the condition $\varepsilon_L
= \varepsilon_R = \mu$ ($E_{e,L} + U_1 = E_{e,R} + U_0 = \mu$)
(see the inset of Fig. A.2(a)). Consequently, there is a very weak
electron-hole asymmetry near $\mu$ caused by $\Gamma_{e,L}\neq
\Gamma_{e,R}$, resulting in an almost negligible electron heat
current rectification shown in Fig. A.2(c).

To further analyze the effects of electron hole asymmetry, we
present $N_{j,\sigma}$, $V_{\text{th}}$, and $Q_e$ for
$\varepsilon_R = \mu +10$ meV in diagrams (d), (e), and (f). As
shown in Fig. A.2(e), the maximum $V_{\text{th}}$ is significantly
enhanced by at least two orders of magnitude. Moreover,
$V_{\text{th}}$ exhibits an ${\cal N}$-shape behavior. The
electron heat rectification behavior is observed when
$\varepsilon_R \neq \varepsilon_L = \mu$. Based on the results
presented in Fig. A.2, it can be concluded that electron heat
current rectification favors ${\cal
T}^{2-site}_{LR(RL)}(\varepsilon)$ with a strong electron-hole
asymmetry near $\mu$.

\begin{figure}[h]
\centering
\includegraphics[angle=0,scale=0.3]{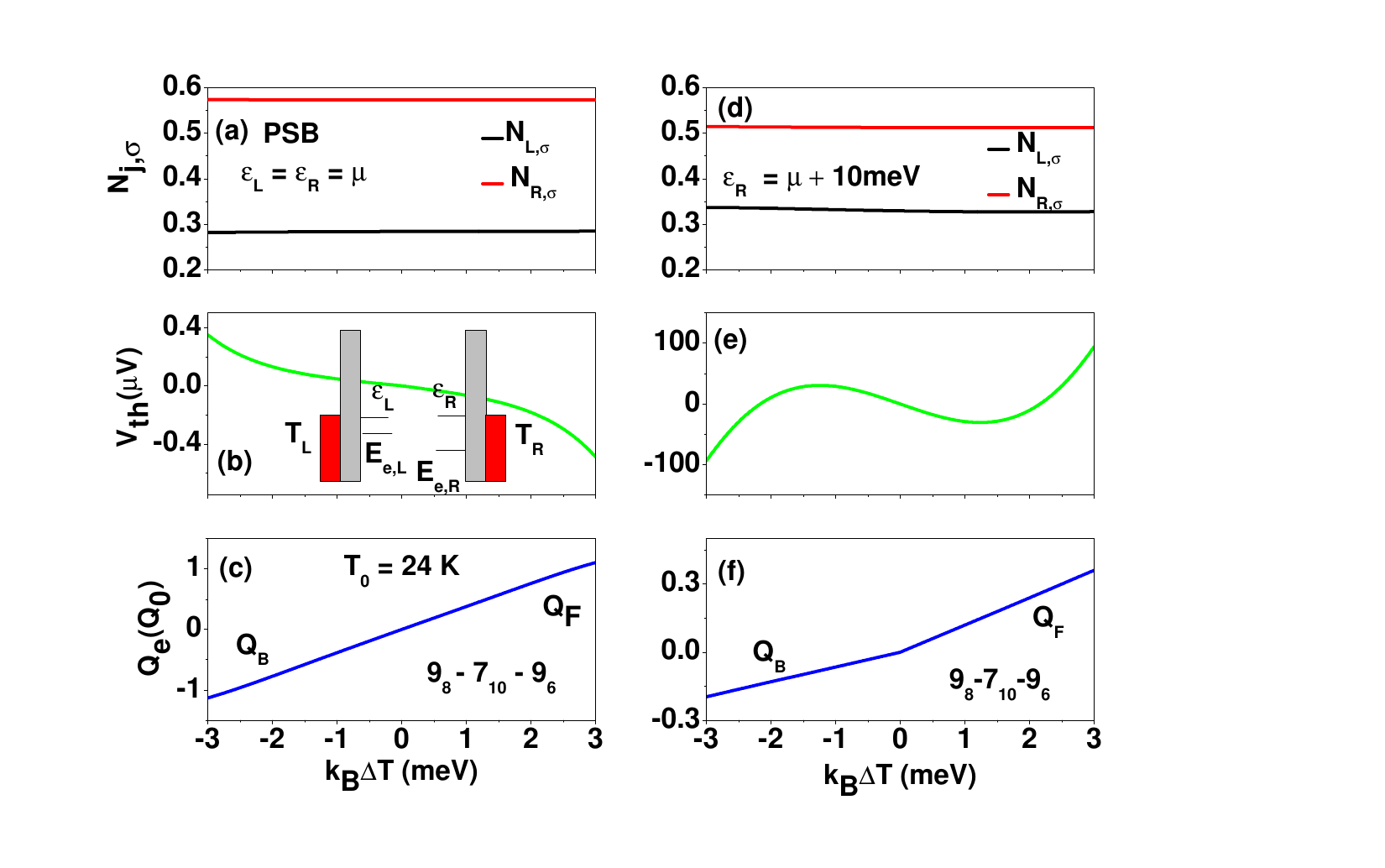}
\caption{(a) Electron occupation numbers ($N_{j,\sigma}$), (b)
thermal voltage ($V_{th}$), and (c) electron heat current ($Q_e$)
as functions of temperature bias $\Delta T$ at $T_0 = 24$ K for
$9_8-7_{10}-9_6$ AGNR heterostructures with Pauli spin blockade
(PSB) configurations, as shown in the inset with $\varepsilon_L =
E_{e,L} + U_1$ and $\varepsilon_R = E_{e,R} + U_0$, satisfying
$\varepsilon_L = \varepsilon_R = \mu$. Diagrams (d), (e) and (f)
display $N_{j,\sigma}$, $V_{th}$ and $Q_e$ for $\varepsilon_R
=\mu+10$~meV.The electron heat current is measured in units of
$Q_0 = 77.2$ pW.}
\end{figure}




\setcounter{section}{0}
\setcounter{equation}{0} 

\mbox{}\\





\end{document}